\documentclass[12pt]{iopart}
\usepackage{iopams}
\usepackage{setstack}
\usepackage{graphicx}
\usepackage{epsfig}

\usepackage{epstopdf}
\usepackage{times}
\usepackage{txfonts}
\usepackage{xcolor}

\newcommand{\beq}{\begin{equation}}
\newcommand{\eeq}{\end{equation}}
\newcommand{\bea}{\begin{eqnarray}}
\newcommand{\eea}{\end{eqnarray}}

\newcommand{\<}{\langle}
\renewcommand{\>}{\rangle}

\newcommand{\commentout}[1]{{}}

\begin{document}
\bibliographystyle{h-physrev3}
\title[Quantum dynamics of atomic bright solitons]{Quantum dynamics of atomic bright solitons under splitting and re-collision, and implications for interferometry}
\author{A. D. Martin}
\address{School of Mathematics, University of Southampton, Southampton SO17 1BJ, United Kingdom}
\address{Department of Physics, University of Otago, P. O. Box 56, Dunedin, New Zealand}
\author{J. Ruostekoski}
\address{School of Mathematics, University of Southampton, Southampton SO17 1BJ, United Kingdom}

\date{\today}
\begin{abstract}
We numerically study the classical and quantum dynamics of an atomic bright soliton in a highly-elongated one-dimensional harmonic trap with a Gaussian barrier. In the regime of the recent experiment by Dyke {\it et al.}, the system realizes a coherent nonlinear soliton beam-splitter and interferometer whose accuracy we analyze. In the case of tighter radial trap confinement and enhanced quantum fluctuations, a split soliton can represent a spin-squeezed, or alternatively, a fragmented condensate with reduced phase-coherence that can be measured by interfering the split soliton by the barrier.
We also find large quantum mechanical uncertainties in the soliton's position and momentum due to nonlinear interaction with the barrier.
\end{abstract}

 \maketitle

\section{Introduction}

Bright solitons are solitary waves robust to collisions with each other and with external potentials, and are commonly studied within many physical systems including nonlinear optics \cite{Haus_RMP_1996,Stegeman_Science_1999} and Bose-Einstein condensates (BEC) \cite{Khaykovich_Science_2002,Strecker_Nature_2002,Cornish_PRL_2006,Abdullaev_IntJModPhysB_2005}. In the BECs bright solitons form peaks in atomic density against a negligible background density and are found in condensates of attractively-interacting atoms, or repulsively-interacting atoms in an optical lattice (band gap solitons) \cite{Eiermann_PRL_2004}. In two-species BECs a bright solitary wave can be stabilized in one component while trapped inside a co-propagating dark solitary wave
(phase kink) in the other component, forming a dark-bright vector soliton \cite{Dutton_Science_2001,Busch_PRL_2001,Anderson_PRL_2001,Becker_Nature_2008,Shrestha,Yan}.

Despite the considerable theoretical and experimental efforts, the main research emphasis on bright solitons has been in classical mean-field properties, e.g., within the Gross-Pitaevskii equation (GPE). Soliton dynamics in BEC experiments \cite{Khaykovich_Science_2002, Strecker_Nature_2002, Cornish_PRL_2006} have been adequately described using the GPE, although quantum and thermal fluctuations are suggested as a seed for condensate collapse mechanisms \cite{Buljan_PRL_2005} and soliton-train formation \cite{AlKhawaja_PRL_2002}. Recently, however, it has been suggested that quantum fluctuations cause the fragmentation of a soliton train \cite{Streltsov_PRL_2011,Wuster_NJPhys_2009} and may already be more important in the present experiments than previously thought. Quantum effects in bright soliton dynamics have been studied in a nonlinear optics context \cite{Lau_PRA_1989,Lau_PRA_1989_b,Drummond_JOptSocAmB_1987,werner,Fini_PRA_1998,optsoliton,jenkins},
and also in few atom strongly-interacting systems (far from the regime we consider in this paper), specifically macroscopic superposition (Schr\"{o}dinger cat) states of solitons \cite{Cederbaum_PRL_2008, Weiss_PRL_2009}, tunneling through a barrier in a Bose-Hubbard model \cite{Glick_unpublished_2011}, and soliton entanglement \cite{Lewenstein_NewJPhys_2009}. Quantum effects have also been predicted in gap solitons \cite{Ahufinger_PRL_2005,Ahufinger_PRL_2005_b}.

Motivated by the recent bright soliton experiments~\cite{Huletexp} we study the dynamics, reflection, transmission, and splitting of bright solitons in a combined trap of a highly-elongated, prolate harmonic potential and a Gaussian optical barrier. The experimental conditions of Ref.~\cite{Huletexp} correspond to large atom numbers in a one-dimensional trap which is not radially very tightly confined. Quantum fluctuations in the ground state of such a system are weak and we first analyze the classical mean-field dynamics in this limit. Displacing the condensate from the equilibrium position at the trap centre initiates dipolar oscillations of the BEC which forms a propagating soliton. We introduce a Gaussian barrier at the trap centre over which a soliton is either completely transmitted, reflected or split into a transmitted and reflected component. The system realizes a strongly nonlinear, phase-sensitive soliton beam-splitter and interferometer. It has been previously suggested that the robust character of bright solitons lends itself to precision measurements, e.g., of surface reflectivity \cite{Cornish_PhysicaD_2009} and for atom interferometry \cite{Parker_JPhysB_2008}.

In order to analyze the atom number statistics of the soliton interferometer formed by the laser barrier acting as a beam-splitter, we extend the theoretical study beyond the classical mean-field approach by studying the soliton dynamics in the stochastic phase-space representation within the truncated Wigner approximation (TWA) \cite{Drummond_EPhysLett_1993,Steel_PRA_1998,Sinatra_JPhysB_2002,Isella_PRA_2006,Blakie_AdvInPhys_2008,Martin_PRL_2010,POL10}. Although the output of the highly nonlinear beam-splitter is very sensitive to the relative phase difference between the solitons suggesting a possibility for high precision interferometry, the numerical analysis of the atom statistics in the TWA simulations shows that the nonlinear soliton-barrier interactions enhance fluctuations of the relative soliton atom numbers in the interferometer output above those given by the classical limit of binomial statistics. This enhancement of fluctuations is large enough to cancel any benefit from the improved sensitivity to the phase values and happens even in the limit of negligible phase decoherence.

We also show that, in a radially more tightly-confined trap and with smaller atom numbers, quantum fluctuations can be significantly enhanced, resulting in dissipative soliton dynamics.
From the TWA simulations we calculate ensemble averages that provide quantum mechanical expectation values and uncertainties of the soliton's position and momentum, as well as the relative atom number statistics and phase-coherence of a soliton pair. Moreover, individual stochastic realizations of the TWA dynamics represent possible outcomes of single experimental runs, revealing considerable variations within an ensemble, e.g., with one trajectory displaying transmission of a soliton through the barrier and another one reflection. We also find that, due to the interaction with the barrier, when the soliton's kinetic energy dominates the nonlinear interaction energy the soliton may split into a spin-squeezed soliton pair propagating to opposite directions. For enhanced fluctuations the initial condensate undergoes fragmentation into a mutually incoherent soliton pair. We show that the loss of phase-coherence in the condensate fragmentation can be measured by studying atom number fluctuations in the recombination process of the solitons by the barrier. Finally, we also propose an approach that could potentially exploit quantum phenomena in the spin-squeezing of the solitons and overcome the shot-noise limit of interferometers based on classical atomic samples.

\section{Soliton Experiments}\label{Sec_Expt}

Experiments on atomic BECs with attractively interacting atoms have produced single or multiple bright solitons, using Li-7 \cite{Strecker_Nature_2002, Khaykovich_Science_2002, Huletexp}, or Rb-85 \cite{Cornish_PRL_2006} atoms trapped in a cigar-shaped geometry. In all of these experiments, a Feshbach resonance was used to tune the s-wave scattering length of the inter-particle interactions to the desired negative value, producing a controllable attractive interaction between the atoms. A stable condensate takes the form of a single soliton \cite{Khaykovich_Science_2002,Huletexp}. Otherwise, if the attractive interactions are too large to be stabilized by the quantum pressure of the trapped atoms, the condensate collapses, leaving behind a predictable number of solitons which may each oscillate within the trap and collide with each other many times during an experimental run \cite{Cornish_PRL_2006}.
In the experiments in Refs.~\cite{Strecker_Nature_2002,Cornish_PRL_2006} the stability of multiple solitons was attributed to the relative-phase of $\pi$ between neighbouring solitons that leads to repulsive interactions between the adjacent solitons and the characteristic collective oscillatory dynamics of the solitons \cite{AlKhawaja_PRL_2002}. The essential condition for the well-defined relative phase of $\pi$ to remain throughout the dynamics is that the solitons were assumed to be sufficiently phase-coherent.  The phase-coherence of solitons has been the subject of some previous theoretical investigations \cite{Buljan_PRL_2005}, and in particular it was suggested in Ref.~\cite{Streltsov_PRL_2011} that the loss of phase-coherence in the soliton trains could cause rapid fragmentation of the condensate.

\begin{figure}[tbp]
\centering
\includegraphics[width=\columnwidth,trim = 0mm 0mm 0mm 0.1mm,clip=true]{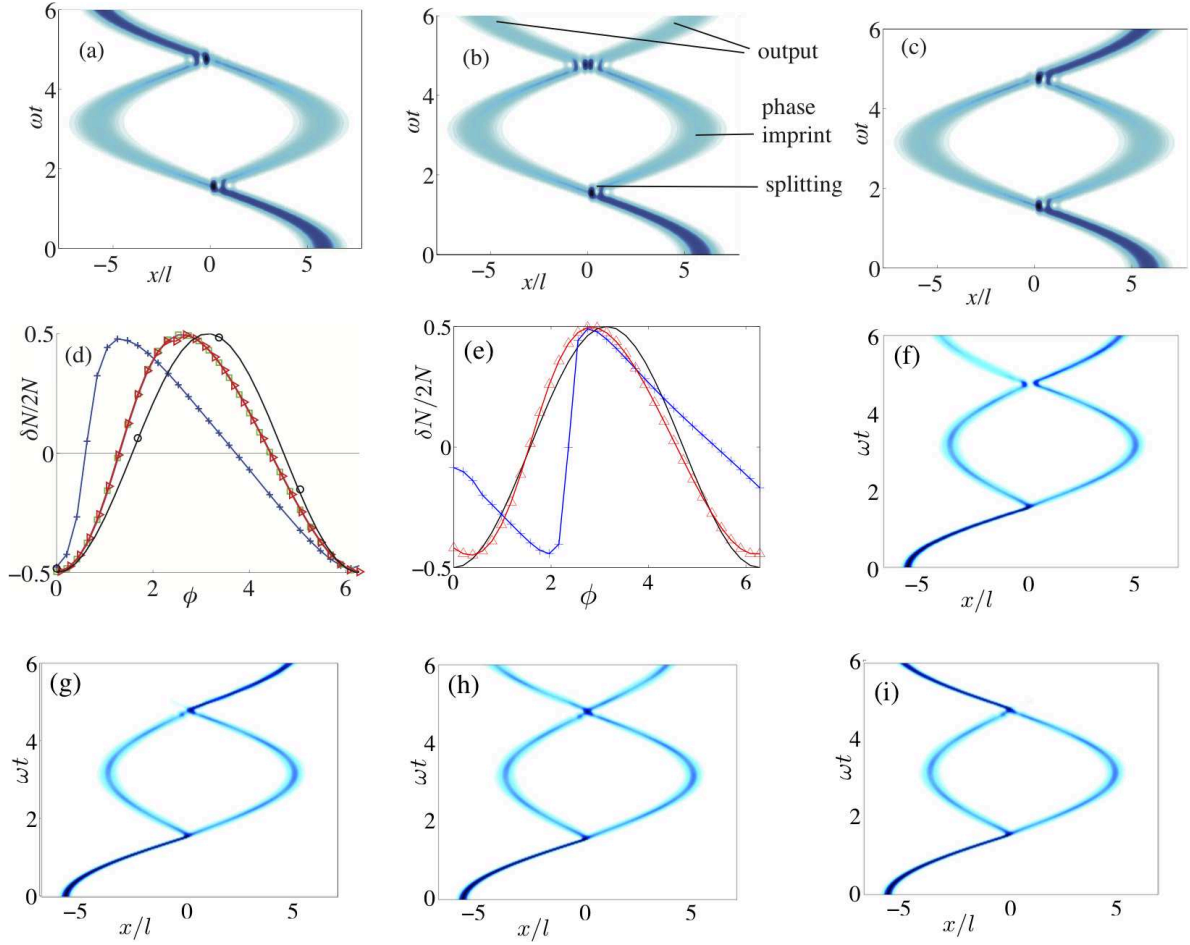}
\caption{(a)-(c), (f)-(i) Atomic density in the 1D harmonic trap as a function of time for a propagating soliton within the classical mean-field analysis in the absence of quantum fluctuations. The soliton is initially displaced from the trap minimum and hits a potential barrier at the trap centre, subsequently splitting into two solitons that propagate into opposite directions. The two solitons then collide at the barrier at the trap centre and, depending on their relative phase, emerge either to the negative $x$ direction, positive $x$ direction, or the both directions. In (a)-(c) the nonlinearity takes the value $|U|=3\hbar \omega$, the barrier width $\sigma=0.021l$ and the barrier heights $V_0$ for the splitting and solitons' re-collision are both close to $144\hbar\omega$. In (f)-(i) $|U|=15\hbar\omega$, $\sigma=0.14l$, the barrier height $V_0\simeq 23\hbar \omega$ for the splitting, and the final barrier height $V_0\simeq24\hbar\omega$. The kinetic energy per atom in the soliton as it collides with the barrier is $T_E\simeq17\hbar\omega$. The right-hand soliton is multiplied by a complex phase: (a) $\phi=0$, (b) $\phi=1.2$ (c) $\phi=2.6$, (f) $\phi=0$, (g) $\phi=2.0$, (h) $\phi=2.4$ and (i) $\phi=2.7$. The different stages of the interferometer scheme are marked on (b). (d) Relative atom number of the outgoing solitons $\delta N$ scaled by twice the total atom number $2N$ as a function of the phase $\phi$ applied by direct multiplication of the soliton wavefunction before the second collision with the barrier (red line with triangles) for nonlinearity $|U|=3\hbar \omega$, and by ramping a constant potential across one of the solitons [see Eqs.\ (\ref{Eq_ImprintFull}) and (\ref{Eq_Imprint})] for $|U|=3\hbar \omega$ (green line with squares) and $|U|=10\hbar \omega$ (blue line with crosses), and for comparison the exact solution $(\cos\phi)/2$ [see Eq.\ (\ref{secondcol})] for $U=0$ (black line) with numerical points from the GPE simulation (black circles) for the corresponding parameters. (e) as (d), but for a thicker barrier with $\sigma=0.14l$. The trajectories in (a)-(c) have phases chosen such that at the end of the dynamics, all the atoms populate the left/right outgoing soliton or are evenly distributed between both outgoing solitons.  The trajectory in (f) has the imprinted phase $\phi=0$, and the trajectories in (g)-(i) span the region in (e) with large gradient, and have imprinted phases as described above.} \label{Fig_ClassicalDensity}
\end{figure}

The recent experiment \cite{Huletexp}, which forms the starting point of our theoretical study, is an ideal system to study the soliton phase-coherence.
In this experiment, a BEC comprising a single soliton of $2\times 10^5$ $^7$Li atoms
was formed in the centre of an elongated 1D harmonic trap that was generated by a tightly-focused laser beam.
The BEC was made weakly attractive by tuning the scattering length to $a\simeq -0.5 a_0$ via a Feshbach resonance, where $a_0$ denotes the Bohr radius. The radial and axial angular frequencies were $\omega_r=2\pi\times 300$Hz and $\omega=2\pi\times 3$Hz, respectively, such that although the system was not strictly 1D, it was stabilized against collapse, with the chemical potential $\mu \lesssim \hbar \omega_r$, resulting in significantly reduced density fluctuations along the radial direction.
The potential minimum along the axial direction was suddenly displaced such that the soliton was placed in a nonequilibrium position, inducing dipolar oscillations for the condensate in the trap. A thin barrier, generated by a near-resonant cylindrically focused laser beam was inserted at the centre of the trap. The propagating soliton interacted with the barrier, and depending on its kinetic energy and the size of the barrier, underwent a reflection, transmission, or was split into two.
The split solitons propagated in opposite directions before re-colliding at the barrier where they were observed to output in either direction [demonstrated in the numerical simulations in Fig.\ \ref{Fig_ClassicalDensity}(a)-(c)].
In Secs.~\ref{Sec_Classical} and \ref{Sec_Quantum} we investigate the splitting behavior and the phase-dependence of the solitons' re-collision at the barrier first using classical mean-field analysis and then going beyond the mean field by including atom statistics.

The soliton-barrier system provides an interesting atom-optical realization of a coherent, nonlinear soliton beam-splitter and interferometer. The nonlinear classical dynamics in this system can also lead to the formation of a bound pair of solitons or a `soliton molecule' as a result of the interaction of a soliton with the barrier \cite{AlKhawaja_unpublished_2010}. An alternative method to split a single soliton coherently into two by means of applying density modulation with resonant pulses between different internal states was proposed in Ref.~\cite{Billam_PRA_2011}.
\section{Classical Dynamics} \label{Sec_Classical}
We first investigate the classical periodic dynamics and splitting of a soliton in a harmonic potential within the GPE. The system may be considered to be in a 1D regime when the chemical potential $\mu$ and thermal energy $k_BT$ are small compared with the transverse trap energy $\hbar\omega_r$. For simplicity, here we assume that the system is sufficiently tightly confined in the radial direction so that the radial degrees of freedom are confined to a Gaussian radial ground-state wavefunction, and the dynamics are described by the 1D GPE:
\begin{equation}
i\hbar \frac{\partial}{\partial t}\psi=-\frac{\hbar^2}{2m}\frac{\partial^2}{\partial x^2}\psi+ g|\psi|^{2}\psi+V_{ext}\psi,\label{Eq_GPE1}
\end{equation}
where $g=2\hbar\omega_r a$, $\omega_r$ is the angular frequency of the radial trapping potential, $V_{ext}$ is the external axial potential including the axial harmonic potential $V=m\omega^2 x^2/2$, the Gaussian barrier [Eq.\ (\ref{Eq_Barrier})] and, in some simulations, a phase imprinting potential  [Eq.\ (\ref{Eq_Imprint})].

In the absence of the external trapping potential, the 1D GPE with attractive interactions ($g<0$) supports soliton solutions with the centre-of-mass position for the soliton $j$ denoted by $q_j$ and its centre-of-mass velocity $\nu_j=d q_j/d t$.
Using the notation $\bar x =x/l_r, \bar q_j = q_j/l_r,\bar a=a/l_r, \tau= t\omega_r, \bar \nu_j=d \bar q_j/d \tau$, where $l_r=(\hbar/m\omega_r)^{1/2}$, the soliton solution may be written in the following form  \cite{Zakharov_JETP_1972,Zakharov_JETP_1972_b}:
\beq
{\psi}_{j}(x,t)= \sqrt{\frac{|\bar a|N_{j}^2}{2 l_r}} \mathrm{sech} \left[ |\bar a| N_j(\bar x-\bar q_{j})\right] e^{i {\bar v_{j}(\bar x-\bar q_{j})}} e^{i \phi_j(\tau)},
\label{Sol_GPE2}
\eeq
with
\begin{eqnarray}
\phi_j(\tau) &={N_j^2 \bar a^2}\tau+S_j(\tau)+\phi^{(0)}_j, \label{Eq_classicalphase}\\
S_j(\tau) &=\tau {\bar \nu_j^2}/2.\label{ActionHomo}
\end{eqnarray}
Here $N_{j}$ is the number of atoms in the $j$th soliton, and $\sum_j N_j=N$, where $N$ is the total atom number in the system. The spatially independent part of each soliton's phase has been encapsulated in $\phi_j(\tau)$ and $\phi^{(0)}_j$ denotes the initial value of the phase. For the case of a single initial soliton, the initial phase value $\phi^{(0)}_j$ represents the global phase of the wavefunction and is inconsequential to the dynamics. If several solitons are present, they may approach each other, collide, and then leave each other unscathed up to shifts in position and phase.
In the harmonically trapped system we consider, we calculate the ground-state soliton profile by imaginary-time propagation of the GPE, and find it to be slightly distorted compared with the homogeneous solution. In the presence of a harmonic trap the dynamics is no longer integrable, but it is easy to show that in a weak harmonic potential, for the approximate solution of a displaced soliton,  $q_j$ and $v_j$ oscillate harmonically, such that
\begin{eqnarray}
\bar{q}_j=&\bar{q}_{j0}\cos\left(\bar{\omega}\tau\right)+\bar{v}_{j0}\sin\left(\bar{\omega}\tau\right)\label{Eq_sho}\\
\bar{v}_j=&\bar{v}_{j0}\cos \left(\bar{\omega}\tau \right)+\bar{q}_{j0}\sin \left(\bar{\omega}\tau \right)
\end{eqnarray}
where $q_{j0}$ and $v_{j0}$ are the initial position and velocity of the $j$th soliton, $\bar{\omega}=\omega/\omega_r$, and the term,
\begin{equation}
S_j(\tau)=\left(\frac{\bar{v}_{j0}^2}{2\bar{\omega}}-\frac{\bar{q}_{j0}^2\bar{\omega}}{2}\right) \frac{\sin\left(2\bar{\omega}\tau\right)}{2}+\frac{\bar{q}_{j0}\bar{v}_{j0}}{2} \cos\left(2\bar{\omega}\tau\right)-\frac{\bar{q}_{j0}\bar{v}_{j0}}{2}, \label{Action}
\end{equation}
replaces Eq.\ (\ref{ActionHomo}) in the phase dynamics of Eqs.~(\ref{Eq_classicalphase}) and~(\ref{Sol_GPE2}) \cite{Martin_PRA_2008}. From the soliton wavefunction solution given by Eq.\ (\ref{Sol_GPE2}) we can calculate the expectation values for the first and second powers of soliton position and momentum as
\begin{eqnarray}
\<x_{j}\>_s &=  \int x |\psi_j(x)|^2 \epsilon_j^{-1}dx =q_j,\label{Eq_Avx}\\
\<x^2_{j}\>_s &= \int x^2 |\psi_j(x)|^2 \epsilon_j^{-1} dx =\frac{l_r^4\pi^2}{12a^2 N_j^2}+q_j^2,\\
\<p_{j}\>_s &=-i\hbar \int  \psi_j(x)^*\frac{\partial{\psi_j(x)}}{\partial x} \epsilon_j^{-1}dx =mv_{j},\\
\<p^2_{j}\>_s &=-\hbar^2  \int  \psi_j(x)^*\frac{\partial^{2}{\psi_j(x)}}{\partial x^{2}} \epsilon_j^{-1}dx =\frac{\hbar^2 a^2N_j^2}{3 l_r^4}+m^2v_{j}^{2},
\end{eqnarray}
where \begin{equation}
\epsilon_{j} \equiv \int_{-\infty}^{\infty}|\psi_{j}(x)|^2 dx=N_j.\label{Eq_AvN}
\end{equation}
Here, the notation $\<...\>_s$ indicates we are averaging across one soliton wavefunction. These relations provide
the corresponding position and momentum uncertainties of a single soliton in the classical GPE mean-field solutions
\begin{eqnarray}
(\Delta x_j)_s &=\left(\langle {x_j}^2 \rangle_s-\langle {x_j} \rangle_s^2\right)^{1/2}= \frac{l_r^2\pi}{2 \sqrt{3}a N_j}\,,\label{xwidth}\\
(\Delta p_j)_s &=\left(\langle {p_j}^2 \rangle_s-\langle {p_j} \rangle_s^2\right)^{1/2}=\frac{\hbar a N_j}{\sqrt{3} l_r^2}\label{pwidth}\,.
\end{eqnarray}
These uncertainties are determined by the width of the soliton's wavefunction in Eq.\ (\ref{Sol_GPE2}).

We model the experiment \cite{Huletexp} in which a single soliton was split into two solitons of approximately equal size (in this paper, by soliton size we refer to the number of atoms in the soliton).
We assume the atoms are initially in equilibrium in the centre of the trapping potential and are described by Eqs.~(\ref{Sol_GPE2}) and~(\ref{Eq_sho}-\ref{Action}). We numerically simulate the displacement of the soliton by a distance $x_d$ (equivalent to the experimental trap displacement of $-x_d$), followed by the application of a Gaussian barrier of width $\sigma$ and height $V_0$,
{\begin{equation}
V_{b}(x)=V_0 \exp\left(-{x^2\over \sigma^2}\right),\label{Eq_Barrier}
\end{equation}
at the trap centre.  The soliton starts propagating from the edge of the trap towards the barrier. We investigate the behavior of the colliding soliton by varying the ratio between the nonlinearity $|U|$ and the centre-of-mass kinetic energy per atom, $T_E$, for the soliton as it hits the barrier, defined by
\begin{equation}
U=\frac{gN}{l},\quad  l=\sqrt{\frac{\hbar}{m\omega}}\,,\label{Eq_U}
\end{equation}
\begin{equation}
T_E=\frac{1}{2}m\omega^2x_d^2\,.\label{eq:TE}
\end{equation}
Here $l$ denotes the axial harmonic oscillator length and $x_d$ the initial centre-of-mass displacement of the soliton from the trap centre. The expression (\ref{eq:TE}) represents the initial potential energy when the centre-of-mass velocity of the soliton is zero. This potential energy is assumed to be converter to kinetic energy at the trap minimum.

We study the soliton dynamics using parameters close to those of the recent experiment \cite{Huletexp}.
For example, in the experiment the Gaussian laser barrier of a 5$\mu$m waist ($1/e^2$ intensity radius) in the trap with $\omega=2\pi\times 3$Hz corresponds to the width $\sigma\simeq0.16 l$ in Eq.~(\ref{Eq_Barrier}). A typical experimentally studied value for the ratio between the kinetic energy of the soliton and the nonlinearity was $T_E/|U|\simeq 5.5$ (for $|U|\simeq 48 \hbar\omega$ and the initial trap displacement of 0.5mm, or $x_d\simeq 23 l$).
We vary the barrier width between the very narrow values of  $\sigma\simeq0.012 l$ or $\sigma\simeq0.021 l$ and broad values with $\sigma\simeq0.14 l$ or, in Sec.~\ref{Sec_pq}, with $\sigma\simeq0.32 l$. For the narrow barrier case the interaction time of the soliton with the potential is minimized, representing a sharp ideal barrier, reminiscent of a $\delta$-function potential. Here $\sigma\simeq0.14 l$ is close to the experimental values of Ref.~\cite{Huletexp}.
For $^7$Li atoms in the trap of Ref.~\cite{Huletexp} with $\omega=2\pi\times 3$Hz, the narrow barrier width of $\sigma\simeq0.021 l$ can be achieved with a diffraction-limited focusing of a laser with the wavelength 650nm or shorter. In a weaker trap the barrier may be experimentally realized with a less tightly-focused laser.
In most simulations we vary the nonlinearity and use the initial displacement $x_d\simeq 5.9l$, such that the kinetic energy as the soliton hits the barrier $T_E\simeq 17\hbar\omega$. For the nonlinearities $|U|\simeq 3\hbar\omega$ this yields approximately the same ratio $T_E/|U|\simeq 5.5$ as in Ref.~\cite{Huletexp}. For the largest nonlinearity we examine ($|U|=15 \hbar \omega$) we also consider smaller displacements $x_d=3.2l$ such that $T_E\simeq 5.2\hbar\omega$.

When the nonlinear energy dominates (we examine the case $|U|=15 \hbar \omega$ and $T_E\simeq 5.2\hbar\omega$) the soliton does not split but may become stuck to the barrier for a finite time, and/or be either completely transmitted or completely reflected gaining a $\pi/2$ phase-shift on reflection [i.e., the reflected soliton is obtained by replacing $\phi^{(0)}_j$ by $ \phi^{(0)}_j +\pi/2$ and $v_{j}$ by $-v_j$ in Eq.~(\ref{Sol_GPE2})].
The $\pi/2$ phase shift may first appear surprising, but it is analogous to the phase difference between a reflected and transmitted  electromagnetic wave from a finite-width dielectric slab enclosed between two optically identical media \cite{Efimov_JOptA_2001}.

Lowering $|U|$ or increasing the initial displacement (as in the example case we studied with $T_E\simeq 17\hbar\omega$) the soliton may split into two solitons (a transmitted part and a reflected part, the latter of which gains a $\pi/2$ phase-shift), as weaker attractive interactions are less inclined to hold the soliton together. The proportion of the reflected and transmitted component depends on the interactions, as well as the barrier parameters (see Fig. \ref{Fig_ClassicalSplit}). We define the relative atom number $\delta N$ and relative phase $\delta\phi$ between the solitons by
\begin{eqnarray}
\delta N&=N_2-N_1,\\
\delta \phi&=\phi_2-\phi_1,
\end{eqnarray}
where $N_2$ and $\phi_2$ are the number of atoms and the phase [see Eqs.~(\ref{Sol_GPE2}) and~(\ref{Eq_classicalphase})] in the left soliton ($x<0$), and $N_1$ and $\phi_1$ are the number of atoms and the phase in the right soliton ($x>0$).

We display the splitting dynamics as a function of the interactions and the barrier height and width [$V_0$ and $\sigma$ in Eq.\ (\ref{Eq_Barrier})] in Fig.~\ref{Fig_ClassicalSplit}.
In the specific parameter regime shown in Fig.~\ref{Fig_ClassicalSplit}(a), for a given barrier height a larger proportion of soliton is transmitted for small $|U|$ than for large $|U|$. Consequently, a larger barrier is required to split a soliton into two equal parts when the interactions are reduced. Figure \ref{Fig_ClassicalSplit}(b) demonstrates the intuitive result that a barrier of a smaller width has to be higher to split a soliton into two equal parts.
\begin{figure}[tbp]
\centering
\includegraphics[width=\columnwidth,trim = 0mm 0mm 0mm 0.1mm,clip=true]{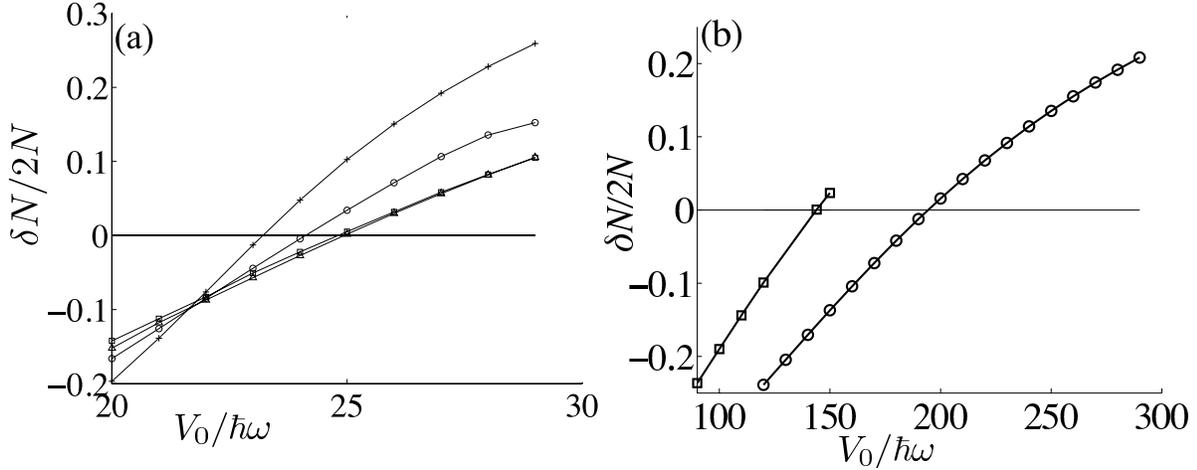}
\caption{(a) Relative atom number between two solitons $\delta N$ in the classical mean-field dynamics. The two solitons are formed by splitting a single soliton into transmitted and reflected components by a potential barrier, defined in Eq.\ (\ref{Eq_Barrier}). The relative atom number is normalized by twice the total atom number $2N$ and is displayed as a function of the potential barrier height $V_0$. The width of the barrier is $\sigma=0.14l$ and the kinetic energy per atom of the soliton as it hits the barrier is $T_E\simeq17\hbar\omega$. The nonlinearity (occurring in Eq.\ (\ref{Eq_GPE1})) is  $|U|=3\hbar \omega$ (squares), 5$\hbar \omega$ (triangles) 10$\hbar \omega$ (circles) and $15 \hbar \omega$ (crosses). (b) as (a), but with $|U|=3\hbar \omega$, and with a barrier of width $\sigma\simeq0.021l$ (squares) and $\sigma\simeq0.012l$ (circles).} \label{Fig_ClassicalSplit}
\end{figure}
%

After a soliton is split into two solitons of equal size, which subsequently evolve in the harmonic trap, the solitons re-collide at the barrier with a relative phase of $\delta\phi=\pi/2$. At this point, the solitons are half the size of the original soliton and will interact in the same manner but as if the nonlinearity $|U|$ were half the strength. To eliminate this effect, we reduce the barrier height between the splitting of the first soliton and the subsequent collision of the solitons at the barrier to a height that would split a soliton of size $N/2$ into two equal parts.
If the solitons undergo free evolution in the harmonic trap between the splitting and recombination, they usually recombine to form a single soliton [see Fig.~\ref{Fig_ClassicalDensity}(a)]. We also find, however, an exceptional case, where the nonlinearity is large, and the barrier is sufficiently wide for a prolonged nonlinear interaction with the soliton [e.g., the trajectory in Fig.\ref{Fig_ClassicalDensity}(f) with parameters $\sigma=0.14l$ and $|U|=15\hbar\omega$]. In this case, the solitons do not fully combine, but two solitons emerge from the collision at the barrier. Moreover, in the case of wide barriers a soliton can collide inelastically, such that even for a barrier chosen to split a soliton into two parts of equal atom number the system can be asymmetric. In this case the reflected soliton is slower than the transmitted soliton, and performs lower amplitude oscillations [see Fig.\ \ref{Fig_ClassicalDensity}(f)-(i)].

We study the relationship between the relative phase $\delta\phi$ and relative atom number $\delta N$ in the outgoing solitons by applying a constant relative phase difference $\phi$ by directly multiplying the wavefunction of one of the solitons by the phase factor $\exp\left[i \phi\right]$, such that its wavefunction $\psi_j$ becomes $\psi_j\exp\left[i \phi\right]$. We find that generally, the solitons do not fully recombine, but there are two outgoing solitons of different sizes with $\delta N$ that depends on the imprinted phase. Hence, we can use the system as an interferometer in detecting weak forces that accumulate a phase difference between the solitons.  Since the two split solitons are spatially separated, in principle they can experience a difference in some weak force, e.g., in gravity that generates the change in the relative phase between the solitons. The accumulated relative phase difference can be measured by observing the relative atom number between the reflected and transmitted parts after the collision with the barrier.
In order to simulate such a phase imprinting procedure which may be performed in an experiment we consider applying an example potential of the form
\begin{equation}
V=V_{r}(t)\Theta(x),\label{Eq_ImprintFull}
\end{equation}
where
\begin{equation}
V_{r}(t)=V_m \exp\left[-\beta^2(t-t_m)^2\right],\label{Eq_Imprint}
\end{equation}
between times $t=t_i$ and $t_f$, where $t_m=(t_f+t_i)/2$, and $\Theta(x)$ is the Heaviside step function, so the potential is felt by one soliton and not the other; classically, in the absence of any phase fluctuations, this potential imprints the phase
\beq
\phi=\int_{t_i}^{t_f}dt {V_{r}\over \hbar}=\sqrt{\pi}{V_m\over 2\hbar\beta}\left[\mbox{Erf}(\beta t_f)-\mbox{Erf}(\beta t_i)\right]\,,
\eeq where Erf is the error function defined by:
\begin{equation}
\mbox{Erf}(z)=\frac{2}{\sqrt{\pi}}\int_0^z\exp\left[-y^2\right] dy\,.
\end{equation}
We choose $\beta=\omega$ and show in Fig.\ \ref{Fig_ClassicalDensity}(d) that the phase imprinting by the example potential yields almost identical results to multiplying one soliton wavefunction by $\exp\left[i \phi\right]$.

The dependence of the relative atom number on the relative phase between the two solitons is easiest to understand in the case of a non-interacting gas (i.e., $U=0$). We assume that in the first collision the barrier splits a soliton of containing $N$ atoms into two equal parts and in the second collision the barrier is changed so that it splits a soliton containing $N/2$ atoms into two equal parts. If we assume that the phase $\phi$ is imprinted on the initially reflected soliton, that soliton has then acquired a phase difference $-\delta\phi=\phi+\pi/2$ with respect to the initially transmitted soliton. After the second collision with the barrier, the soliton propagating to the negative $x$ direction consists of a transmitted component of the initially reflected soliton and the reflected component of the initially transmitted soliton (which now acquires a $\pi/2$ phase shift upon the reflection). Adding both contributions together results in the atom number $N \cos^2(\phi/2)$ for the soliton propagating to the negative $x$ direction, where $N$ denotes the atom number of the initial soliton. The atom number of the soliton propagating to the positive $x$ direction is $N \sin^2(\phi/2)$ and
the relative atom number after the second collision with the barrier reads
\beq
{\delta N\over N}=\cos(\phi)\,.\label{secondcol}
\eeq

We find that in the presence of nonlinear interactions the dependence of the splitting on the relative phase differs from the noninteracting result. Moreover, the splitting behaviour is different for different barrier widths. If the solitons are split and recombined by a narrow barrier (we simulate a barrier of width $\sigma\simeq0.021l$), the dependence of the relative atom number in the output solitons on the phase difference is qualitatively similar to the noninteracting case, but the sensitivity to the imprinted phase is enhanced. As shown in Fig.~\ref{Fig_ClassicalDensity}(d), when the nonlinearity $|U|$ is increased, the functional dependence deviates from the exact sinusoidal behaviour of the non-interacting case due to the increasingly nonlinear interaction with the barrier. As $|U|$ increases, the value of $\phi=\phi_m$ at which $\delta N/N$ is a maximum is reduced from $\phi=\pi$ at $U=0$. Consequently, the gradient of $\delta N/N$ becomes steeper for $\phi<\phi_m$ and shallower for $\phi>\phi_m$. Further narrowing the barrier has only a small effect on this deviation: using a barrier of approximately half the width (reduced from $\sigma\simeq0.021l$ to $\sigma\simeq0.012l$)  and height $V_0\simeq200\hbar\omega$ (chosen to split the soliton into two equal parts) only reduces the maximum difference in $\delta N/N$ between the $|U|=3\hbar\omega$ and $U=0$ cases by 3\%.
For a wider barrier (we simulate a barrier of width $\sigma=0.14l$ that is close to the experimental values of Ref.~\cite{Huletexp}), at weak nonlinearities, e.g., $|U|=3 \hbar\omega$, the deviation from the noninteracting result is similar to that of a narrow barrier case. However, as the nonlinearity is increased, e.g, to $|U|=15\hbar\omega$, the curve $[\delta N/N](\phi)$ becomes strongly distorted from a sinusoid. We also find that the first minimum in $\delta N/N$ is shifted from $\phi=0$ to $\phi\simeq 2$, as shown in Fig.\ \ref{Fig_ClassicalDensity}(e). The significant asymmetry of the curve representing the relative atom number as a function of the imprinted phase is reminiscent of the nonlinear behaviour of bistable systems. This could indicate the existence of more than one solution of $\delta N$ for a given value of $\phi$.

The steep gradient of $\delta N/N$ as a function of the applied phase $\phi$ for the nonlinear solitons indicates that measuring the relative atom number between the solitons exiting the barrier could potentially provide a very accurate detection method for the applied phase and, hence, an accurate weak force detection. The classical analysis of the mean-field dynamics based on the GPE simulations, however, does not yield any information about the statistical fluctuations of the atom numbers, resulting from the nonlinear interactions of the soliton with the barrier. Additionally, quantum fluctuations may increase phase  fluctuations between the atoms, reducing the accuracy of the detection of the applied force that induces the relative phase $\phi$. In order to analyze the effect of statistical atom number fluctuations in the collisions of the solitons with the barrier, we simulate the soliton dynamics by incorporating the atom statistics in TWA. In Section \ref{Sec_Quantum_Inter} we show how the nonlinear soliton-barrier interaction increases the uncertainty in the relative atom number between the solitons and effectively cancels any potential benefit that could be obtained from the sensitivity of $\delta N/N$ to variations of the phase $\phi$.

\section{Truncated Wigner Approximation}\label{Sec_TWA}
In this section we outline the methods we employ for extending our study to include quantum fluctuations. A common approach in stochastic phase-space methods is to unravel the evolution dynamics into stochastic trajectories, each of which obeys the classical mean-field dynamics. Each trajectory is a representative of a probabilistic initial state distribution that is numerically generated by stochastic sampling. The probability distribution is selected so that one can synthesize as closely as necessary the quantum statistical correlations or thermal distribution of the atoms in the initial thermal equilibrium state. The phase-space representation that most accurately reproduces classical mean-field dynamics is the Wigner representation because of the `correct amount' of quantum noise in the initial state  \cite{Gardiner_Book,Sinatra_JPhysB_2002}.

Here we introduce stochastic phase-space methods for the atom dynamics within TWA \cite{Drummond_EPhysLett_1993,Steel_PRA_1998,Sinatra_JPhysB_2002,Isella_PRA_2006,Blakie_AdvInPhys_2008,Martin_PRL_2010,POL10} in order to investigate atom number statistics of the solitons, collapse of relative phase-coherence, and the condensate fragmentation. In our approach we follow the formalism of Ref.~\cite{Isella_PRA_2006}. We consider a unitary evolution of the 1D atomic gas without additional couplings to environment and replace quantum field operators $(\hat{\psi},\hat{\psi}^{\dagger})$ by the classical fields $(\psi_{W},\psi^{*}_{W})$ that are the
stochastic phase-space representation of the full field operator. The nonlinear dynamics of each stochastic trajectory, sampled from the distribution of the initial conditions (we sample between 100 and 1000 stochastic realizations of $\psi_W$ for each parameter regime), is then governed by the GPE evolution [Eq.\ (\ref{Eq_GPE1})]. The quantum fluctuations are stochastically included in the initial state $\psi_{W}(x,t=0)$. We solve the initial state within the Bogoliubov approximation when the full quantum field operator is given by
\begin{equation}
\hat{\psi}(x,0)=\psi_0\hat{\alpha_0}+\sum_j \left[u_j(x)\hat{\alpha}_j -v_j^*(x)\hat{\alpha}_j^{\dagger}\right]\,.
\end{equation}
Here the phonon mode functions $u_j$ and $v_j$ are calculated numerically. The stochastic initial state is obtained by replacing the phonon mode operators $\hat{\alpha}_j,\hat{\alpha}_j^{\dagger}$ by classical stochastic variables \cite{Isella_PRA_2006}. Within TWA, ensemble averages and variances of observables correspond to quantum mechanical expectation values and uncertainties. Individual realizations of $\psi_W$ represent outcomes of individual experimental runs. The TWA method becomes more accurate when the modes are highly populated. However, the TWA simulations have been successful in providing qualitative descriptions of experiments \cite{Fertig_PRL_2005} on dissipative quantum dynamics in 1D lattices even for significant ground state depletion and at very small atom numbers, as shown in Ref.~\cite{Isella_PRL_2005}.

Although the 1D dynamic model neglects density fluctuations in the radial direction, it can also provide a good qualitative description of several experimental systems in which the radial confinement is not strong. This was demonstrated in Ref.~\cite{Gross_PRA_2011} where experimentally observed atom number squeezing and reduced on-site atom number fluctuations were compared with numerical 1D TWA simulation results. It has been suggested, however, that in a condensate collapse mechanism and soliton train formation a full 3D treatment of radially weakly confined systems can lead to distinct features in dynamics \cite{Wuster_NJPhys_2009}.

In order to calculate quantum mechanical expectation values and correlation functions for the soliton observables, we follow a method similar to that in Refs.~\cite{Isella_PRA_2006,Isella_PRA_2005} and project the stochastic field $\psi_W(x,t)$ onto some particular set of mode functions, here represented by the soliton solutions $\Upsilon_j(x,t)$. The amplitude of the $j$th soliton reads
\begin{equation}
a_j(t)=\int dx\, \Upsilon^{*}_j(x,t) \psi_W(x,t), \label{Eq_Projection}
\end{equation}
which corresponds to the stochastic Wigner representation for the annihilation operator of the soliton $j$. We define the normalized nonlinear mode functions $\Upsilon_j(x,t)$ in such a way that the amplitudes $a_j(t)$ incorporate information about the atom number and the phase of the soliton $j$
\beq
{\Upsilon}_{j}(x,t) \equiv {e^{-i\phi_j(\tau)}\over \sqrt{N_j}}\, \psi_j(x,t)
=\sqrt{\frac{|\bar a|N_j}{2 l_r}} \mathrm{sech} \left[ |\bar a| N_j(\bar x-\bar q_{j})\right] e^{i {\bar v_{j}(\bar x-\bar q_{j})}},
\label{Mode_function}
\eeq
where $\psi_j(x,t)$ is given by Eq.~(\ref{Sol_GPE2}), and $\bar{x}$, $\bar{a}$, $\bar{q}_j$ and $\bar{v}_j$ and $l_r$ are the variables defined in Sec.~\ref{Sec_Classical}. The phase of each soliton in each stochastic realization of the TWA dynamics is given by the complex phase of $a_j$. In the classical limit, this phase corresponds to $\phi_j(\tau)$ in Eq.~(\ref{Eq_classicalphase}).

In order to track the mode functions $\Upsilon_j(x,t)$ in time,
we numerically calculate the soliton position in each realization of $\psi_W$ by integrating the centre-of-mass position $\int x|\psi_W|^2dx$ one or both sides of the barrier (if one or two solitons
are present) and identify this with the soliton position(s), $q_j$. We calculate the time derivative of $q_j$ along the each trajectory in $\psi_W$ and identify this as the velocity $v_j$. The number of atoms $N_j$ in each soliton is evaluated by integrating $|\psi_W|^2$ around the soliton's centre-of-mass position.

Quantum mechanical expectation values and correlation functions related to the soliton's atom number and the phase-coherence may then be calculated in the TWA simulations from the ensemble averages consisting of products of $a_j(t)$'s. The relative phase-coherence between two bright solitons $ j=1,2$ at times between the soliton splitting and recombination are obtained from the normalized first order correlation function
\begin{equation}
c=\frac{|\langle \hat{a}_1^{\dagger}\hat{a}_2 \rangle|}{\sqrt{\langle \hat{N}_1\rangle\langle \hat{N}_2\rangle}}, \label{Eq_coherence}
\end{equation}%
where the expectation value of the number of atoms in the $j$th soliton is obtained from
\beq
\langle \hat{N}_j\rangle= \langle a_j^{*} a_j\rangle_W - 1/2\,.
\eeq
The expectation value for the relative atom number $\delta \hat{N}$ and its fluctuations are given by
\begin{eqnarray}
\langle \delta \hat{N}\rangle & = \langle\hat N_2\rangle - \langle \hat N_1\rangle=\langle a_2^{*} a_2\rangle_W - \langle a_1^{*} a_1\rangle_W\,,\\
\left(\Delta \delta N\right)^{2} & = \langle\left( \hat{N}_{2}-\hat{N}_{1}\right)^2\rangle-\langle \hat{N}_{2}-\hat{N}_{1}\rangle^2 \nonumber\\
&=\langle \left(a_{2}^{*}a_{2}-a_{1}^{*}a_{1}\right)^{2} \rangle_{W}-\langle a_{2}^{*}a_{2}-a_{1}^{*}a_{1}\rangle_{W}^{2}-1/2. \label{Eq_deltaJ}
\end{eqnarray}
Within the Wigner representation, the stochastic variables correspond to symmetrically ordered expectation values, here denoted by $\langle\cdots\rangle_W$, that are obtained in the simulations as ensemble averages. The 1/2 term subtracted from the stochastic expectation value arises from the conversion between symmetrically and normally ordered operator averages. In our simulations typical occupation numbers for the solitons vary between 50 and 400000 atoms.

\section{Quantum dynamics} \label{Sec_Quantum}

We interpolate between regimes of significant quantum fluctuations and the classical mean-field limit, holding the nonlinearity $|U|$ [given by Eq.\ (\ref{Eq_U})] constant and varying $g/N$ ($g$ can be adjusted in experiments, e.g., by varying the transverse confinement or the scattering length using a Feshbach resonance).
In repulsively interacting atomic gases, $g/N$ is proportional to the effective interaction strength $\gamma=m g/\hbar^2 n$, where $n$ denotes the 1D atom density \cite{Kherunstyan_PRA_2005}, that describes the strength of quantum fluctuations in the system. In tightly-confined attractively interacting atomic gases that we consider, $|g/N|$ also indicates the strength of quantum fluctuations.

Fixing $U$ to be constant fixes the number of excited-state atoms depleted from the ground state in the Bogoliubov approximation; hence, varying $g/N$ is tantamount to varying the depleted atom fraction. When the magnitude of $|U|$ is large, i.e., the interactions between atoms are strong and the total atom number is small, quantum fluctuations are enhanced. By increasing $|U|$ we find significant effects on the number fluctuations, phase decoherence and increased soliton position uncertainties for large $|g/N|$. The classical limit is $N\rightarrow\infty$, $g\rightarrow 0$, when the depleted atom fraction becomes negligible. Here the dynamics is generally expected to approach the classical GPE description, the system becomes phase coherent, and the atom number statistics Poissonian.
In the case of strongly nonlinear soliton-barrier interaction, however, we show that the nonlinear dynamics is very sensitive to any small perturbations and the relative atom number statistics of the split solitons considerably exceed the classical shot noise in the limit $N\rightarrow\infty$, $g\rightarrow 0$.
\subsection{Condensate fragmentation, phase-coherence and relative atom number statistics}\label{Sec_Quantum_NPhi}

In Sec.~\ref{Sec_Expt} we discussed the importance of the relative phase-coherence between the solitons in the condensate collapse experiments leading to formation of soliton trains \cite{Strecker_Nature_2002,Khaykovich_Science_2002,Cornish_PRL_2006} and in the experimental realization of an atom-optical soliton beam-splitter and interferometer that require a high level of coherence \cite{Huletexp}. Here we study the effects of quantum fluctuations on the soliton phase-coherence and show how the coherence has important implications for the many-body entanglement in the system, and to possible quantum-enhanced precision measurements. We calculate the relative phase-coherence between the two solitons that are spatially separated by the barrier. We determine how well the phase-coherence is maintained during the spatial separation before the solitons are recombined and under which conditions the phase-coherence is lost and the two solitons represent the fragmentation of the initially phase-coherent condensate.

Quantum fluctuations have been experimentally observed in BEC systems that are divided into spatially separated regions, e.g., by optical lattices \cite{ORZ01,Gerbier_PRL_2006,Li_PRL_2007, Stebby_PRL_2007,Esteve_Nature_2008, Gemelke_Nature_2009, Bakr_Science_2010,Sherson_Nature_2010,Gross_PRA_2011}. The reduced tunneling amplitude for the atoms between adjacent sites and repulsive inter-atomic interactions lead to sub-shot-noise fluctuations in relative atom numbers between different lattice sites, along with reduced relative phase-coherence. Here we find that the interaction of the soliton with the laser barrier results in the splitting of the soliton into a reflected and transmitted soliton that exhibit non-classical relative atom number correlations even in the case of large atom numbers. We also examine how the relative atom number fluctuations between the solitons, when they exit the barrier the second time, are related to the decoherence rate of the solitons and show that the loss of the phase-coherence can be detected by monitoring the atom number statistics of the exiting solitons.

In order to quantify the relative atom number fluctuations between the two split solitons we calculate the relative atom number squeezing $\xi^{2}_{N}$ obtained from
\begin{equation}
\xi^{2}_{N}=\left(\Delta \delta N\right) ^{2}\frac{\langle \hat{N}_{2}\rangle+\langle\hat{N}_{1}\rangle}{4\langle\hat{N}_{2}\rangle\langle\hat{N}_{1}\rangle}.\label{Eq_XiN}
\end{equation}
In the shot-noise limit, the relative atom number fluctuations follow a binomial distribution, and $\xi^{2}_{N}=1$. Interactions between the atoms may suppress the atom number fluctuations such that $\xi^{2}_{N}<1$, in which case the fluctuations are number-squeezed. (When $\langle\hat{N}_{2}\rangle=\langle\hat{N}_{1}\rangle$, this condition is equivalent to $(\Delta \delta N) ^{2}<N$.) Alternatively, the nonlinear character of Eq.\ (\ref{Eq_GPE1}) may cause the system dynamics to be sensitive to small perturbations, such that fluctuations are enhanced above the shot-noise limit ($\xi^{2}_{N}>1$).
We consider how $\xi^{2}_{N}$ depends on the strength of the nonlinearity by varying $U$ [Eq.\ (\ref{Eq_U})] in systems with fixed total atom number $N=1200$ and with centre-of-mass kinetic energy per atom $T_E\simeq17\hbar\omega$ [defined in Eq.\ (\ref{eq:TE})] as the solitons hit the barrier.  We study the time evolution between the first two collisions of the solitons with the barrier, i.e., after a single soliton is split into two and before the first recombination at the barrier. We study barriers of different widths; as in the classical simulations we use both a narrow barrier described by Eq.\ (\ref{Eq_Barrier}) with width $\sigma\simeq0.021l$, which closely represents a sharp ideal barrier, and a wider barrier with width $\sigma=0.14l$ (close to the experimental values of Ref.\ \cite{Huletexp}). Figure \ref{Fig_Coh}(c) shows that for the lowest value of $|U|$ that we consider, the fluctuations reach the shot-noise limit ($\xi^{2}_{N}=1$), as expected for a non-interacting condensate. As the interaction strength is increased to the values $3\hbar\omega \lesssim |U|\lesssim5\hbar\omega $, we find the relative atom number fluctuations to be squeezed, with  $\xi_N^2\simeq 0.89$$(\pm 0.04)$ at $|U|=5\hbar\omega l$  when soliton is split by the narrow barrier ($\sigma=0.021l$). For the wider barrier case ($\sigma=0.14l$), squeezing is weaker, with $\xi_N^2\simeq 0.98(\pm 0.05)$ for the same nonlinearity. We quote the statistical sampling errors in $\xi_N^2$ which occur due to the finite number of stochastic trajectories in each ensemble.
For stronger interactions, the squeezing disappears, and $\xi_N$ becomes much greater than one as the interaction between the soliton and the barrier in the splitting process becomes more nonlinear and sensitive to small perturbations. It is possible that other splitting methods less sensitive to perturbations may produce relative atom number-squeezed solitons in the limit of large $|U|$.
\begin{figure}[tbp]
\centering
\includegraphics[width=\columnwidth,trim = 0mm 0mm 0mm 0.1mm,clip=true]{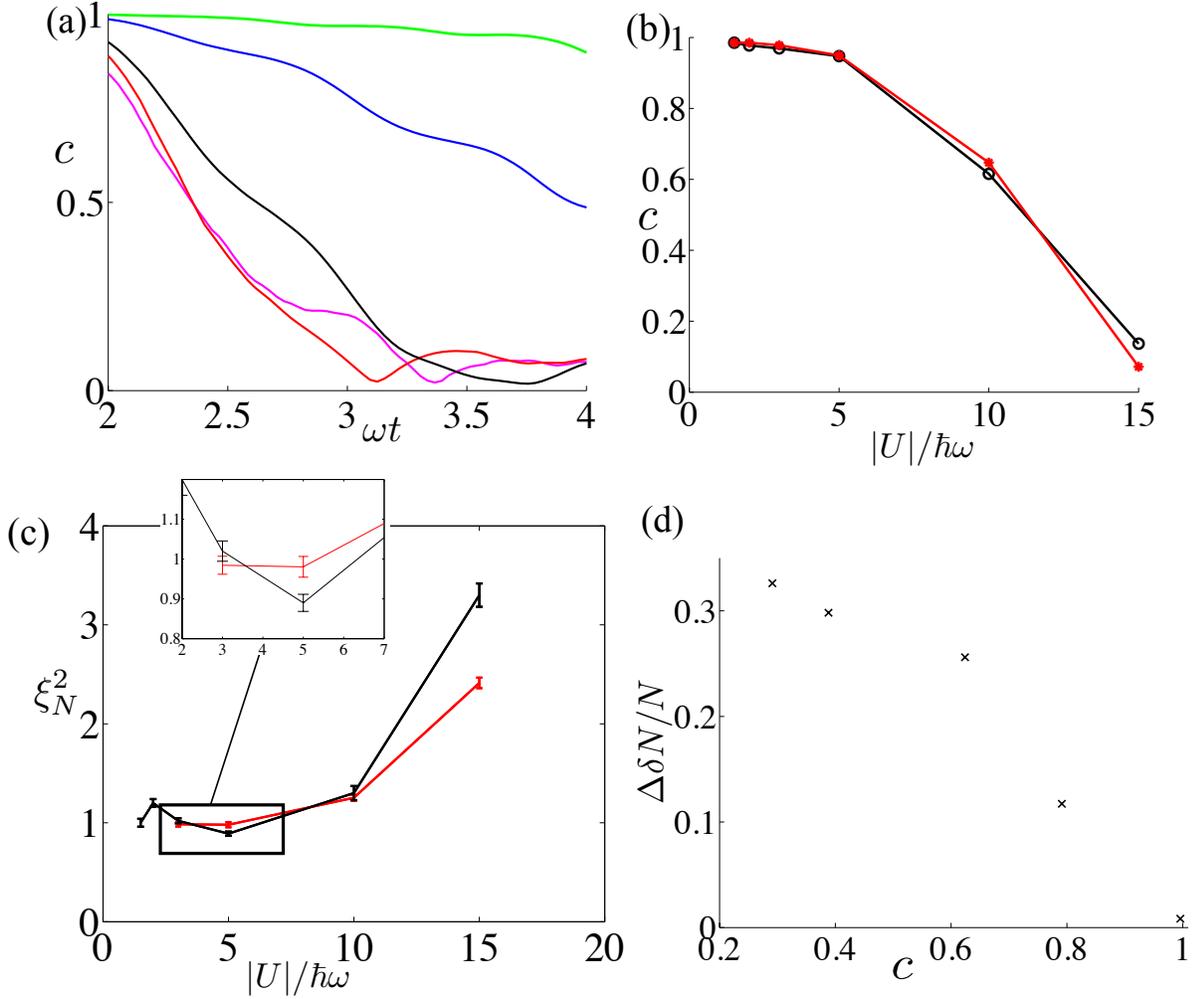}
\caption{(a) Evolution of the relative phase-coherence $c$ between two solitons, between the times $\omega t\simeq2$ (immediately after splitting of the initial soliton by the Gaussian barrier  of the same parameters as in Fig.\ \ref{Fig_ClassicalDensity} (e)-(i)) and $\omega t\simeq4$ (immediately before the solitons re-collide at the barrier). The kinetic energy per atom as the solitons hit the barrier is the same as Fig.\ \ref{Fig_ClassicalDensity} and the nonlinearity $|U|=15\hbar\omega$. The different curves correspond to different values of the fluctuation parameter, corresponding to the atom numbers $N=300$ (magenta), $N=600$ (red), 1200 (black), 8000 (blue) and 80000 (green). The fluctuation parameter $g/N$ varies between approximately $-1.7\times 10^{-4}\hbar\omega l$ and $-2.3\times 10^{-9}\hbar\omega l$. (b) The phase-coherence between solitons, $c$ [given by Eq.\ (\ref{Eq_coherence})], as a function of nonlinearity $|U|$ (in the units of $\hbar\omega$) with $N=1200$ at time $\omega t\simeq 4$, immediately before the solitons re-collide at the barrier. The red line corresponds to a system split by the same barrier as in (a), and the black line to a system split by the narrower barrier used in Fig.\ \ref{Fig_ClassicalDensity}(a)-(d). The barrier height is different for different values of $|U|$ chosen so as to split the soliton in two equal parts. (c) Relative atom number squeezing parameter $\xi_N^2$ [given by Eq.\ (\ref{Eq_XiN})] measured at a time between the splitting of a soliton and re-collision for the same system as (b). The red and black lines are for the same wide and narrow barriers as in (b). The error bars show the statistical uncertainty in the value of $\xi_N^2$ arising due to the finite number of stochastic trajectories in each ensemble. The solitons in the narrow barrier case display stronger relative atom number squeezing and also exhibit spin-squeezing or non-classical entanglement. (d) The quantum uncertainty in the relative atom number between the solitons $\Delta \delta N$, normalized by total atom number $N$ following the solitons' re-collision at the barrier as a function of their relative phase-coherence $c$ at a time immediately before the re-collision, after phase imprinting by a potential given by Eq.\ (\ref{Eq_Imprint}) with amplitude $V_m\simeq3.4\hbar\omega$. The different points are for the same total atom number $N$ but with the nonlinearity $|U|$ varied between the non-interacting and the strongly nonlinear systems.} \label{Fig_Coh}
\end{figure}
%

From the TWA simulations we calculate the dynamics of the relative phase-coherence and the decoherence rate between the two split solitons by evaluating $c$ [Eq.\ (\ref{Eq_coherence})] after the soliton splitting and before the re-collision of the two solitons at the barrier. The value of $c$ close to one indicates a high degree of the relative phase-coherence and the decreasing values of $c$ describe the loss of coherence and the condensate fragmentation. We also evaluate the spin-squeezing between the two solitons, associated with the fictitious pseudo-spin $J=N/2$ of the two-state system of the two solitons. If $\langle \delta \hat{N}\rangle=0$ (e.g., between the soliton splitting and re-collision in the systems we consider), the spin-squeezing parameter is given by
\begin{equation}
\xi^{2}_{S}=\left( \Delta \delta N\right) ^{2}\frac{N}{4\left(\langle \hat{J}_x\rangle^{2}+\langle \hat{J}_{y}\rangle^{2}\right)},
\end{equation}
where $\hat{J}_x =(\hat{a}_2^{\dagger} \hat{a}_1 + \hat{a}_1^{\dagger} \hat{a}_2)/2$ and $\hat{J}_y =i(\hat{a}_2^{\dagger} \hat{a}_1 - \hat{a}_1^{\dagger} \hat{a}_2)/2$. The system is spin-squeezed with the relative atom number representing the squeezed quadrature if $\xi^{2}_{S}<1$, indicating a possibility for quantum-enhanced metrology~\cite{Gross_Nature_2010,Wineland,Holland,Bouyer_PRA_1997,Giovannetti_Science_2004}. The condition  $\xi^{2}_{S}<1$ can also be identified as a signature of quantum many-particle entanglement between the two solitons \cite{Sorensen_Nature_2001}.
When both solitons have large populations, the phase-coherence is reflected in a large value of $\langle \hat{J}_x\rangle^{2}+\langle \hat{J}_{y}\rangle^{2}$, such that   small number fluctuations $\xi_N$ and high phase-coherence are tantamount to spin-squeezing \cite{Esteve_Nature_2008}. Since the relative phase between the reflected and the transmitted solitons $\delta\phi\simeq \pi/2$, the mean value of the pseudo-spin is aligned close to the $y$ axis and $  \xi^{2}_{S} \simeq \xi^{2}_{N}/\<\sin \delta\phi\>^2$, where the phase-coherence $\<\sin \delta\phi\>^2$ could be experimentally measured by interfering the solitons or, as explained below, by monitoring the atom statistics of the output of the soliton beam-splitter after the second collision with the barrier.

We calculate the relative phase-coherence $c$ between the two solitons both in the case of a fixed nonlinearity $|U|$, when the initial quantum depletion is varied by changing the value of $g/N$, and in the case of a fixed total atom number $N$, when we vary the nonlinearity $|U|$. We study the phase-coherence after splitting with the same barrier parameters with which we studied the relative atom number squeezing. The effect of the strength of quantum fluctuations on the decoherence rate of the two solitons is illustrated, for the case of the wide barrier, in Fig.~\ref{Fig_Coh}(a) for $|U|=15\hbar\omega$ (the narrow barrier gives similar results). We show the time evolution of $c$ between the initial splitting and the recombination of the solitons at the barrier, i.e., at the times $2\lesssim \omega t \lesssim 4 $.  The different curves represent different depleted fractions of the atoms from the initial ground state of the single soliton, obtained by varying $g/N$. We find that high relative phase-coherence between the two split solitons is maintained during the entire evolution between the splitting and the recombination at the barrier for $N=80000$ (for this nonlinearity, the fluctuation parameter $g/N\simeq -2.3\times 10^{-11}\hbar\omega l$),
even though the solitons are separated $\sim 10l$ apart. Here $N=80000$  represents almost the classical limit with $c\simeq 1$ and with a very slow decoherence rate. The other curves demonstrate a stronger decoherence rate and the condensate fragmentation, representing a transition from the classical to quantum limit as quantum fluctuations become progressively more prominent. In the case of strong quantum fluctuations ($N=600$, such that $g/N\simeq -4.2\times 10^{-5}\hbar\omega l$) the coherence is already notably reduced immediately after the splitting ($\omega t\simeq 2$) and decays rapidly.
Moreover, we find that the decoherence rate is sensitive to the nonlinearity; for nonlinearities $|U|\lesssim 5\hbar\omega $  the solitons remain phase-coherent with $c\gtrsim0.9$ even for large $g/N$ (atom numbers as low as 600).
We fit the curves in Fig.~\ref{Fig_Coh}(a) (where the wide barrier was used to split the solitons) to the function $c(t)=A \exp[-\Gamma (\omega t)^{\alpha}]$ and obtain $\Gamma\simeq 1.6$, 1.2, 0.22, and 0.03; $\alpha\simeq 1.1$, 1.6, 1.6 and 1.3; $A\simeq 0.85$, 0.89, 0.97 and 0.99 for the atom numbers $N=300$, 1200, 8000, and 80000, respectively. For curves obtained using the narrow barrier, the corresponding values are $\Gamma\simeq 3.1$, 1.5, 0.27, and 0.03; $\alpha\simeq 2.1$, 1.3, 1.4 and 1.3; $A\simeq 0.78$, 0.89, 0.97 and 0.99. Here the fluctuation parameter $g/N$ takes values between $\-1.7\times 10^{-4}\hbar\omega l$ and $-2.3\times 10^{-9}\hbar\omega l$.

We show in Fig.~\ref{Fig_Coh}(b) the relative phase-coherence $c$ for a fixed total atom number $N=1200$, and kinetic energy per atom as the soliton collides with the barrier $T_E\simeq17\hbar\omega$, when we vary the nonlinearity $|U|$. The coherence is evaluated at $\omega t\simeq 4$ -- immediately before the solitons re-collide at the barrier -- providing information about the coherence of the atoms at the soliton beam-splitter. Strong phase-coherence is retained for nonlinearities as strong as $|U| \simeq 5\hbar\omega$. When $|U|$ is increased, the phase-coherence decreases monotonically and with a steep slope for $|U| \gtrsim 5 \hbar\omega$ (e.g., for the atom number $N=1200$, $c\lesssim0.2$ for $|U|=15\hbar\omega$). We find that the behaviour of the coherence $c$ is quantitatively similar for both barrier widths. We fit the curve in Fig.~\ref{Fig_Coh}(b) corresponding to the narrow barrier to the function $c(t)=A\exp[-\Gamma (|U|/\hbar\omega)^{\alpha}]$ and obtain $\Gamma\simeq 1.5\times 10^{-4}$, $\alpha \simeq 3.5$ and $A \simeq 0.98$. At $|U| \simeq 5\hbar\omega$ the relative atom number fluctuations are most squeezed with  $\xi_N^2\simeq 0.89$}$(\pm 0.04)$  when split by the narrow barrier, and $\xi_N^2\simeq 0.98(\pm 0.05)$ when split by the wide barrier. Due to the high relative phase-coherence between the solitons we find that they also exhibit non-classical quantum correlations, or many-body quantum entanglement, as the system is also spin-squeezed when the narrow barrier is used, with  $\xi_S^2\simeq0.95(\pm0.04)$. However, for the wider barrier case the spin-squeezing is lost, with $\xi_S^2\simeq1.08(\pm0.05)$. The example demonstrates how spin-squeezed solitons can be generated even with large atom numbers, providing a very different approach from the optical lattice experiments \cite{Esteve_Nature_2008,Gross_PRA_2011} for producing spin-squeezed atomic samples.

The loss of the relative phase-coherence between the two solitons could be experimentally measured by interfering the solitons. Here we propose an alternative method of experimentally measuring the loss of the relative phase-coherence that is directly based on detecting the fluctuations in the outputs of the soliton beam-splitter.  In the classical mean-field calculations we showed in Sec.~\ref{Sec_Classical} that the relative number of atoms in the solitons after the recombination at the barrier is dependent upon their relative phase, with the non-interacting limit given by Eq.~(\ref{secondcol}). By including quantum fluctuations in the stochastic TWA simulations we demonstrate here that the loss of phase-coherence is correlated with increased relative atom number fluctuations between the two solitons after the second collision with the barrier. Detecting the fluctuations in the relative soliton atom number could therefore provide a direct signature of the relative phase-coherence and the decoherence rate between the solitons.

\begin{figure}[tbp]
\centering
\includegraphics[width=\columnwidth,trim = 0mm 0mm 0mm 0.1mm,clip=true]{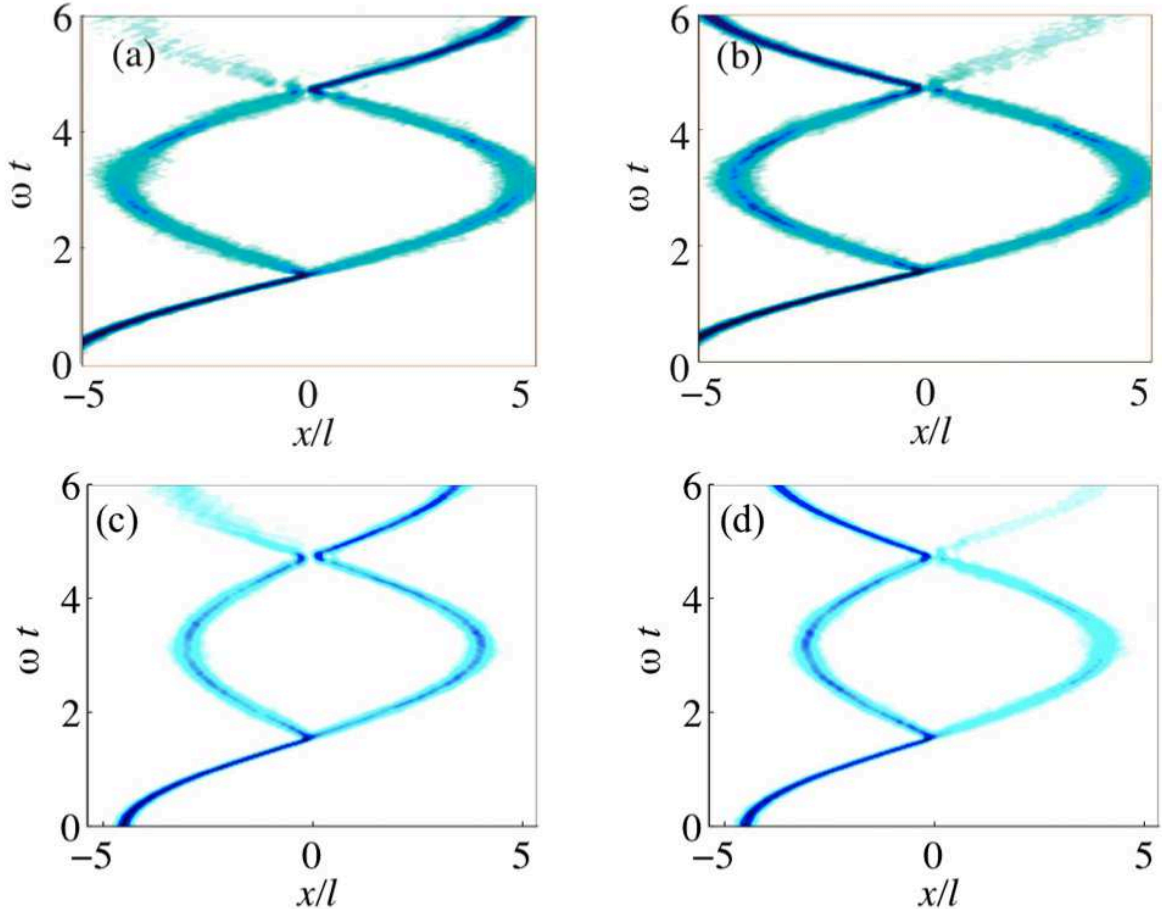}
\caption{Individual stochastic realizations of the Wigner density $|\psi_W|^2$ within the TWA representing possible outcomes of individual experimental runs of the soliton-barrier collision dynamics for mean atom number $N=600$ and nonlinearity $|U|=15\hbar\omega$, with (a)-(b) a narrow barrier of the same dimensions as in Fig.\ \ref{Fig_ClassicalDensity}(a)-(d), and in (c)-(d) a wider barrier of the same dimensions as in Fig.\ \ref{Fig_ClassicalDensity}(e)-(i).  Although the physical parameters in (a)-(b) and also in (c)-(d) are equal, the trajectories correspond to the situations where the solitons exit in the opposite directions from the barrier due to quantum fluctuations.}\label{Fig_Wig2}
\end{figure}

The large relative atom number fluctuations between the two solitons exiting the barrier are illustrated by the particular case shown in Fig.~\ref{Fig_Wig2}. The relative phase-coherence between the solitons before they recombine at the barrier is in this case low with $c\lesssim0.2$. We display in Fig.~\ref{Fig_Wig2}(a)-(b), and (c)-(d) two different examples of two individual stochastic realizations of $\psi_W$ with the same system parameters with the most of the atoms propagating in one realization to the negative $x$ direction and in the other one to the positive $x$ direction, indicating very large atom number fluctuations between different stochastic trajectories. The dependence of the relative atom number fluctuations after the second collision with the barrier on the phase decoherence is characterized more quantitatively by calculating quantum fluctuations from the ensemble averages of the stochastic simulations. We imprint the relative phase of $\delta\phi\simeq \pi/2$ between the solitons as in Sec.~\ref{Sec_Classical}, such that in the classical limit approximately two equal-sized solitons emerge from the barrier [and not only a single recombined soliton; see the non-interacting limit given by Eq.~(\ref{secondcol})]. As shown in Fig.~\ref{Fig_Coh}(d), the loss of the relative phase-coherence between the solitons is clearly signaled in an increase in the relative atom number fluctuations $\Delta\delta N/N$ after the second collision with the barrier. This provides a method of detecting the condensate fragmentation during the soliton splitting by the barrier.

\begin{figure}[tbp]
\centering
\includegraphics[width=\columnwidth]{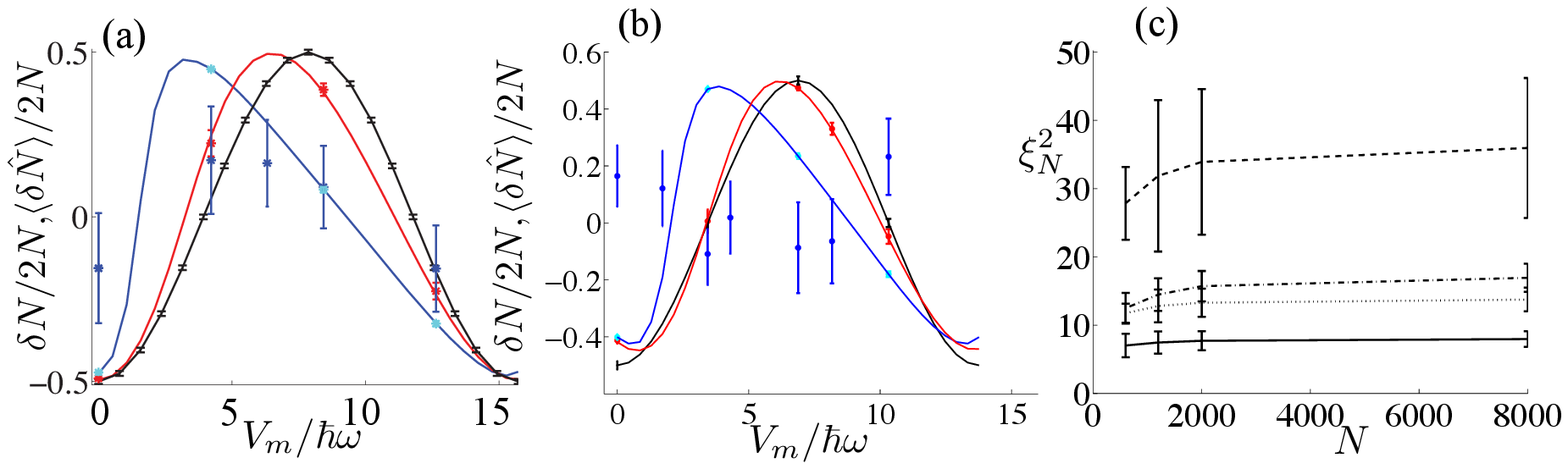}
\caption{(a)-(b) Relative atom number between outgoing solitons $\delta N$ normalized by twice the total atom number $2N$ after collision at the barrier in the soliton recombination following a phase-imprinting by a potential given by Eq.\ (\ref{Eq_Imprint}) with maximum value $V_m$ (in the units of $\hbar\omega$). In (a) the barrier is narrow, with the same width as in Fig.\ref{Fig_ClassicalDensity}(a)-(d), and in (b) the barrier has the same width as in Fig.\ \ref{Fig_ClassicalDensity}(e)-(i).
The lines are from GPE simulations: the nonlinearity takes the values: $|U|=0$ (black line), $3\hbar\omega$ (red line) and $10\hbar\omega$ (blue line). The points and error bars of corresponding colours indicate the quantum mechanical expectation values and uncertainties $\Delta \delta N/2N$ for the TWA simulations of the same nonlinearity and total atom number $N=1200$. Note that the data for $|U|=10\hbar\omega$ do not sit on a curve.  Additional points with cyan error bars are for nonlinearity $|U|=10\hbar\omega$ and total atom number $N=800000$, which sit on the curves produced in the GPE. (c) Relative atom number squeezing parameter $\xi_N^2$ as a function of total atom number $N$ for outgoing solitons. The applied phases are imprinted by the potential Eq.\ (\ref{Eq_Imprint}) with $V_m=0$ (full line), $V_m\simeq3.4\hbar\omega$ (dashed line), 6.9$\hbar\omega$ (dot-dashed line) and 10$\hbar\omega$ (dotted line). The barrier width is the same as in (b), and the nonlinearity $|U|=3\hbar\omega$, such that the system has the parameters of the red curve in (b).} \label{Fig_Jx}
\end{figure}

We show in Fig.~\ref{Fig_Jx}(a)-(b) the quantum mechanical expectation value of the relative atom number $\langle \delta \hat{N}\rangle/2N$ in the solitons exiting the barrier (after the second collision of the atoms with the barrier in the recombination of the solitons) for different nonlinearities $gN$ and total atom number $N$, with the error bars displaying the corresponding quantum uncertainty $(\Delta \delta N)/2N$. We also show $\xi_N^2$ in Fig.~\ref{Fig_Jx}(c) demonstrating how the relative atom number fluctuations typically exceed the shot-noise limit by an order of magnitude. As well as coinciding with increased fluctuations in the relative atom number of the outgoing solitons, the loss of phase-coherence also coincides with a shift in the expectation values $\langle \delta \hat{N}\rangle/2$ towards zero, since due to the negative curvature $ \delta N$ as a function of $\delta\phi$, a symmetric distribution of phases $\delta \phi$ in different stochastic realizations within TWA always lead to a smaller value of $\langle \delta \hat{N}\rangle$. This is illustrated by the points with large error bars in Fig.~\ref{Fig_Jx}(a)-(b) which lie along a flatter profile than the curves for smaller quantum fluctuations.

\subsection{Quantum Bright Soliton Interferometer}\label{Sec_Quantum_Inter}

In Sec.\ \ref{Sec_Classical} we described the classical behavior of a soliton interferometer where during their spatial separation, the solitons can experience a relative phase produced by a small force, e.g., gravity. Such a scheme can be implemented by aligning the axial direction of our trap orthogonal to the direction of the earth's gravitational field, such that the solitons are separated by a vertical displacement $\delta x$ between times $t_i$ and $t_f$. The solitons experience a potential difference of amplitude $V_m=m\tilde{g}\delta x$ which classically imprints the phase $
\phi \sim m\tilde{g}\delta x (t_f-t_i)/\hbar$, where $\tilde{g}$ is the local acceleration due to gravity. Outside the classical regime, this phase will be subject to quantum fluctuations resulting in the phase decoherence described in Sec.\ \ref{Sec_Quantum_NPhi}, providing one limit on the interferometric measurement. The accuracy will be limited also by the relative atom number fluctuations, described by the relative atom number squeezing parameter $\xi_N^2$ in Eq.\ (\ref{Eq_XiN}) which occur due to the sensitivity of the soliton-barrier collision to small  fluctuations. In Sec.~\ref{Sec_Classical} we showed steepening of the gradient of the curve $\delta N/N$ with increasing nonlinearities $|U|$ for applied phases $\phi$ below a maximum value. Here we show any benefit to the sensitivity of the phase measurement due to this steep gradient is effectively cancelled by increased number fluctuations in the outgoing solitons.

In the limit of negligible phase fluctuations, an interferometer's accuracy, $\Delta\phi$, is given in terms of the small phase differences it can resolve. The standard quantum limit for a phase measurement by a classical interferometer with $N$ particles is $\Delta\phi=1/\sqrt{N}$ \cite{Giovannetti_Science_2004}. One may apply a controlled potential to one soliton, such that the relative phase between solitons has some particular expectation value $\langle \delta \hat{\phi}\rangle$ and evaluate the ability of the scheme to distinguish two relative phases: $\langle\delta\hat{\phi}\rangle$ and $\langle \delta\hat{\phi}_1\>$. We consider regimes of high phase-coherence, where the uncertainty is dominated by the enhancement of fluctuations during the collision with the barrier: even for very coherent solitons, $\xi_N^2$ is significant. Thus, we may characterize the accuracy of the interferometer by mapping $\delta N$ as a function of $\langle \delta\hat{\phi} \rangle$ and defining the closest resolvable atom numbers  $\langle \delta\hat{N}\rangle$ and $\langle \delta\hat{N}_1\rangle$ whose separation must be of the order of the fluctuations $|\langle \delta\hat{N}\rangle-\langle \delta\hat{N}_1\rangle |\simeq \Delta \delta N$. We then define the accuracy as $\Delta\phi=\langle\delta\hat{\phi}_1\> -\< \delta\hat{\phi}\rangle$, where $\langle\delta\hat{\phi}\rangle$, $\langle\delta\hat{\phi}_1\rangle$ are the relative phases that would produce the expectations in the relative atom numbers $\langle \delta\hat{N}\rangle$, $\langle\delta\hat{N}_1\rangle$ in the output solitons. We approximately obtain $\Delta\phi\simeq\Delta\delta N/\left[ \partial \langle \delta \hat N\rangle /\partial \<\delta\hat\phi\>\right]$ in which case the accuracy can be improved by having a very sensitive dependence of the relative atom number on the induced relative phase or by suppressed relative atom number fluctuations.

In the non-interacting ($U=0$) case, there is no phase diffusion during the soliton dynamics and no nonlinear interaction between the solitons and the barrier, and the output will be limited by shot noise such that the relative atom number fluctuates according to the classical binomial statistics with $\xi_N^2=1$. For this system we consider a force which in classical GPE simulations would imprint a phase close to $\pi/2$, and within TWA induces the quantum expectation of the relative phase $\langle \delta \phi\rangle=0$ (which contains canceling contributions from the phase shift of the reflected soliton and from the phase imprinted by the potential). Here the gradient of $\delta N$ as a function of $\langle \delta \phi\rangle$ is largest, which should give the best accuracy for the resolvable phase difference $\Delta\phi$. One can evaluate $\Delta\phi$ by mapping the fluctuations in the atom number and the gradient of the curve $\delta N(\phi)$ [e.g., in Fig.~\ref{Fig_Jx}(a)-(b)] to the phase uncertainty by the formula $\Delta\phi\simeq\Delta\delta N/\left[ \partial \langle \delta N\rangle /\partial \delta\phi\right]$. In the non-interacting case this takes the form $\Delta\phi\simeq1/\sqrt{N}$ (the standard quantum limit) in the limit where $N$ is large. 

In Sec.~\ref{Sec_Classical} we showed classical GPE simulation results for interacting solitons that demonstrated high sensitivity of $\delta N$ on $\delta\phi$ with large values of $\partial(\delta N)/\partial(\delta\phi)$. The classical GPE does not provide any information about the atom statistics and if we assume a classical coherent state with $\Delta\delta N\simeq \sqrt{N}$, we would expect an interferometer accuracy that is better than in the ideal gas case or the standard quantum limit. Our stochastic TWA simulations, however, show that $\Delta\delta N$ is substantially increased due to nonlinear interaction of the solitons with the barrier whenever $\partial \langle \delta \hat N\rangle /\partial \<\delta\hat\phi\>$ is large. This happens even in the classical limit of very large atom numbers and negligible phase decoherence with $c\simeq 1$.

In the TWA simulations the phase decoherence is not always negligible and we study the regimes that we found in Sec.~\ref{Sec_Quantum_NPhi} to have strong phase-coherence between the solitons throughout their evolution. This is the case either for weak interactions ($|U|\lesssim 5\hbar\omega $) for all atom numbers we considered $N>600$, or in the classical limit of large $N$ for any $gN$. Within the regimes of high phase-coherence, we consider situations where $\partial \langle \delta \hat N\rangle /\partial \<\delta\hat\phi\>$ is large, corresponding to large classical values of $\partial(\delta N)/\partial(\delta\phi)$ found in Sec.~\ref{Sec_Classical}. (Here the interactions cause the shape of $\left[ \delta N/2N\right]$ to distort, such that its maximum occurs at smaller applied forces, apart from in the exceptional case of a large nonlinearity and wide barrier, where the curve is also shifted with respect to the applied force.) If $\xi_N^2$ were fluctuating according to the classical binomial distribution, the increase in $\partial \langle \delta \hat N\rangle /\partial \<\delta\hat\phi\>$ would decrease the phase uncertainty below the standard quantum limit. However, the interactions also affect the number fluctuations [Fig.\ \ref{Fig_Jx}(a)-(b)] and we find from TWA simulations that when the gradient is steepest, the relative atom number fluctuations are largest, in fact, within these regimes we do not find suppressed fluctuations in the output solitons for any applied force; sensitivity of the re-collision of the solitons at the barrier to small perturbations leads to increased fluctuations in the outgoing solitons for any atom number. For instance, for $|U|=3\hbar\omega$, in Fig.\ \ref{Fig_Jx}(c) we find that for the finite imprinted phases, even in the classical limit of large atom number $N$, for fixed $U$, $\xi_N^2$ converges to approximately 36 for the wider barrier ($\xi_N^2$ converges to 40 for a narrow barrier) at the applied potential for which the number fluctuations have the largest gradient (here $\langle  \delta\hat{N}\rangle/2N\simeq0$). From $\Delta\phi\simeq\Delta\delta N/\left[ \partial \langle \delta \hat N\rangle /\partial \<\delta\hat\phi\>\right]$, we obtain an optimum phase accuracy $\Delta \phi$ of approximately 5 times the standard quantum limit. Interestingly, when no potential is applied ($V_m=0$) and we have $\langle  \delta\hat{N}\rangle\simeq-N$,  for a narrow barrier the fluctuations reflect the initial fluctuations in the total atom number in the classical limit $N\rightarrow \infty$ and $U$ held constant. For a wider barrier, the fluctuations remain finite but smaller than for those in the systems with finite $V_m$. However, in this limit, there is only negligible relative population in the outgoing left soliton, and the relative atom number fluctuations are hard to track numerically. 
We consider the more strongly nonlinear system with $|U|=10\hbar\omega$ and $N=800000$; here there is approximately a three-fold increase in the gradient of $\langle\delta \hat{N}\rangle$ as a function of $\langle\hat{\phi}\rangle$ from the non-interacting case. However, strong number fluctuations (e.g.,  $\xi^2_N\simeq 110$ for the narrow barrier) destroy the ability to measure phase fluctuations below the standard quantum limit despite the increase in the gradient of $\langle\delta \hat{N}\rangle$. Where the gradient of Fig.\ \ref{Fig_Jx}(a) is maximal, this would result in  a phase uncertainty nearly 7 times the standard quantum limit (which would occur if the relative atom number fluctuations were binomial and the gradient was not steepened with respect to the non-interacting case). This phase uncertainty is also 21 times the value which would occur if the system had binomial number fluctuations accompanied by the steep gradient. In simulations extending over multiple periods of the soliton dynamics in the harmonic trap, we find that the number fluctuations increase with repeated soliton collisions with the barrier, which we attribute to the nonlinear interactions between the solitons and the barrier.

\subsection{Sub-shot-noise soliton interferometer}\label{Sec_subshot}

Sub-shot-noise atom number fluctuations provide a central tool for quantum interferometric
sensors that can operate in precision measurements at higher accuracies than the standard quantum limit of
classical atom interferometers \cite{Wineland,Holland,Bouyer_PRA_1997,Giovannetti_Science_2004}. In particular, the phase measurement errors may reach the Heisenberg limit, for the error scaling as $1/N$ with the occupation number $N$, as compared to the standard shot-noise error $1/N^{1/2}$. Reduced atom number fluctuations find interferometric applications in atomic systems \cite{Gross_Nature_2010,Riedel} and the spin-squeezing of a soliton pair (Sec.~\ref{Sec_Quantum}) suggests a possibility for an atomic soliton interferometer realization resulting in quantum statistical properties that can overcome the shot-noise limit of classical atomic samples. Although the soliton interferometer described in the previous section cannot achieve sub-shot-noise accuracies, we propose a scheme that potentially could provide a soliton interferometer overcoming the standard quantum limit that is based on analogous state manipulations as Ref.~\cite{Gross_Nature_2010}. After producing a spin-squeezed soliton pair (Sec.~\ref{Sec_Quantum}), one may perform interferometric manipulation of the internal states of the atoms, instead of letting the solitons collide at the barrier.
Using a $\pi$ pulse one may transfer the atoms in one of the solitons to a different internal state; then, by removing the laser barrier, let the solitons overlap as they reach the centre of the trap. During this spatial overlap one may turn up a Raman coupling between the two internal states for time $\tau$, inducing a rotation in the Bloch sphere by $\exp (i \tau \kappa J_x)$ evolution. In order to minimize the effects of interspecies interactions between the two solitons during the overlap, we require the strength of the Raman coupling $\kappa$ to be sufficiently strong compared to the interaction strength between the two states. By choosing $\tau \kappa=\pi/2$, one can thus transfer number squeezed states to phase squeezed states, such that any accumulated relative phase can be measured with accuracy below classical precision after another $\pi/2$ rotation of the Bloch sphere.

\subsection{Quantum statistics of the soliton's position and momentum}\label{Sec_pq}
\begin{figure}[tbp]
\centering
\includegraphics[width=\columnwidth]{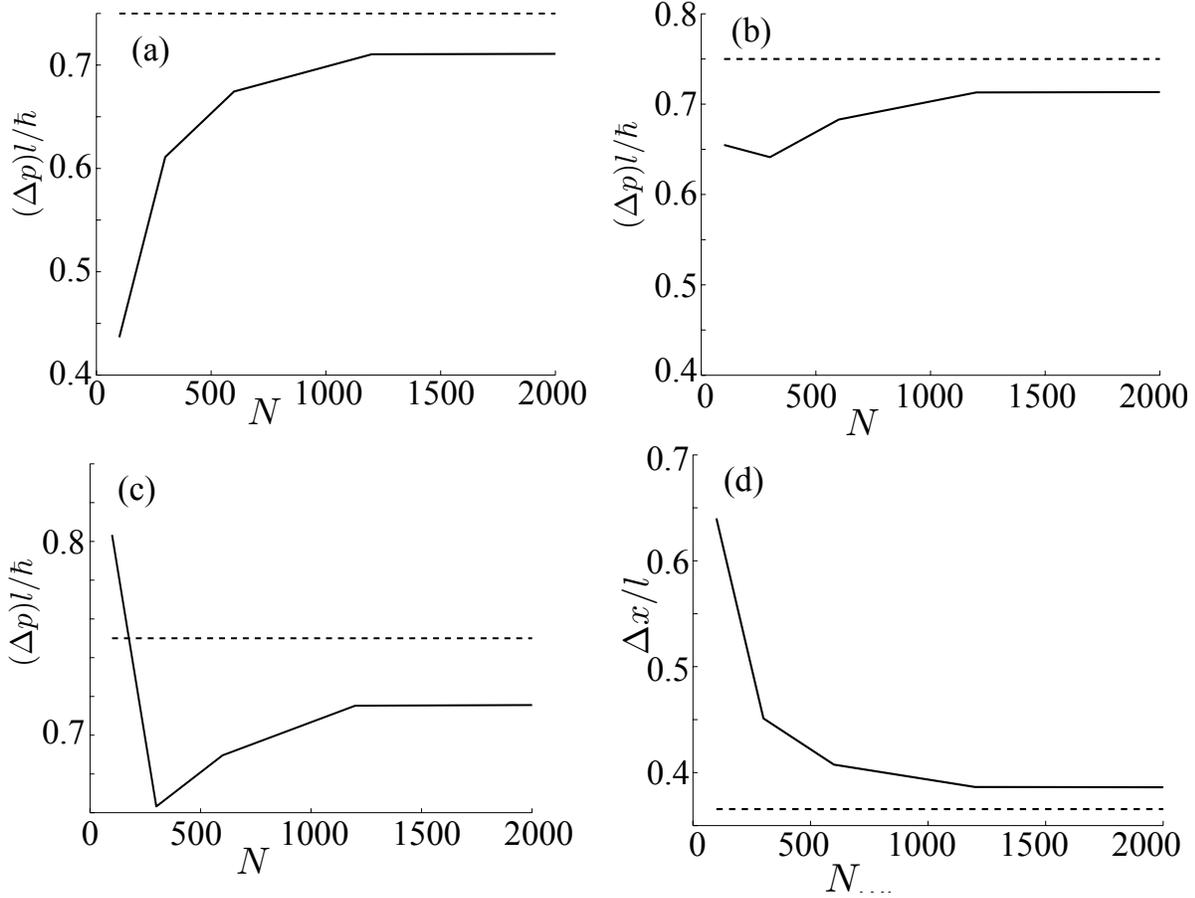}
\caption{(a)-(c) Quantum uncertainty in soliton momentum $\Delta p$ as a function of total atom number $N$ during evolution in a harmonic potential (with no Gaussian barrier included) at times  $\omega t\simeq$ (a) 0.15, (b) 0.70 and (c) 1.3.  (d) Quantum uncertainty in soliton position $\Delta x$ at time $\omega t\simeq0.15$ ($\Delta x$ remains approximately the same at later times). The soliton has kinetic energy per atom as it reaches the trap centre $T_E\simeq17\hbar\omega$ and nonlinearity $|U|=3\hbar\omega$. $\Delta p$ and $\Delta x$ reach their classical (GPE) limits, represented by the dotted lines, before $N=80000$ - well outside the plotting range. We found the statistical errors on these data to be  negligibly small. For example, for $N=600$, at $\omega t\simeq1.3$ the error in $(\Delta p)l/\hbar$ is approximately $1.7\times 10^{-3}$ and the error in $\Delta x/l$ is approximately $2.1\times 10^{-4}$.} \label{Fig_VelocityUncertainty}
\end{figure}	
\begin{figure}[tbp]
\centering
\includegraphics[width=\columnwidth,trim = 0mm 0mm 0mm 0.1mm,clip=true]{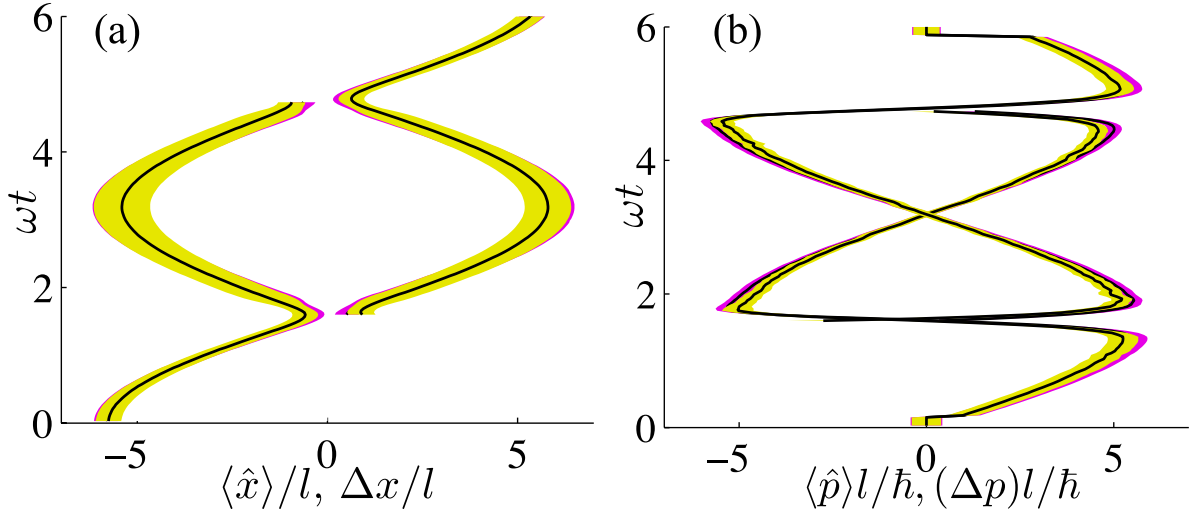}
\caption{Trajectories of the quantum mechanical expectation values and uncertainties in soliton position (a) and momentum (b) for a soliton of the same kinetic energy and split by a barrier of the same parameters as in Fig.\ \ref{Fig_ClassicalDensity} into two solitons which subsequently re-combine at the barrier. The quantum expectation value of the position in (a) and the momentum in (b) are marked by the dark solid lines, whereas the quantum uncertainties are defined by the corresponding shaded areas. The nonlinearity $|U|=3\hbar \omega$ and the total number of atoms $N=1200$ (magenta) and $N=300$ (yellow) such that $g/N=-2.8\times 10^{-6}\hbar\omega l$ and $-3.3\times 10^{-5}\hbar\omega l$ respectively.} \label{Fig_px1}
\end{figure}
\begin{figure}[tbp]
\centering
\includegraphics[width=0.9\columnwidth]{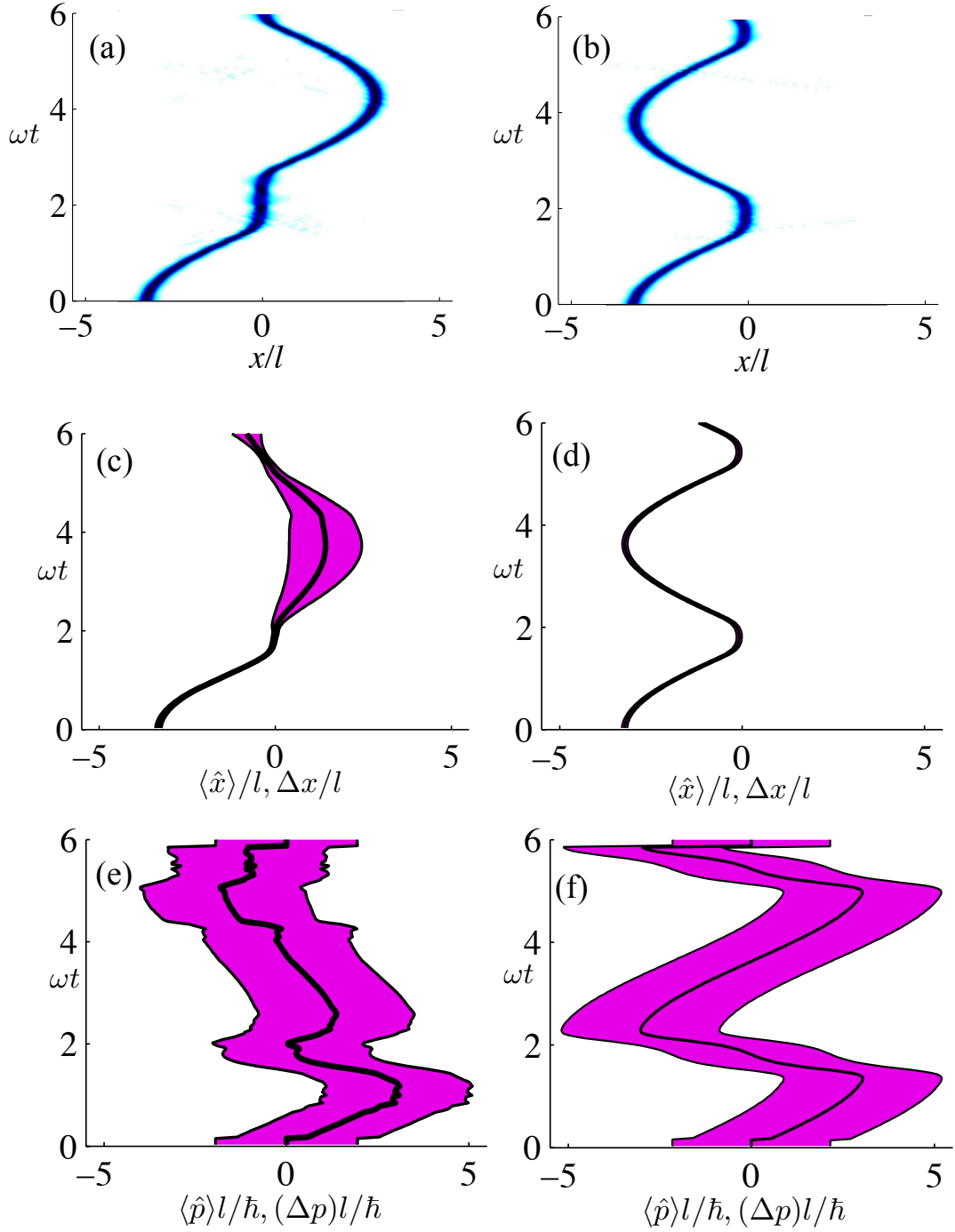}
\caption{(a)-(b) Individual stochastic realizations of the Wigner density $|\psi_W|^2$ within the TWA for nonlinearity $|U|=15\hbar\omega$, in the regimes where solitons travel in different directions after colliding with a barrier described by Eq.\ (\ref{Eq_Barrier}). The barrier height $V_0\simeq6.0 \hbar \omega$ and width $\sigma=0.32l$, the total atom number $N=600$ and the kinetic energy per atom of the soliton as it hits the barrier $T_E\simeq 5.2\hbar\omega$. (c)-(f) Quantum expectation values  and uncertainties in soliton position (c)-(d) and momentum (e)-(f) for the same nonlinearity $U$ and potential barrier as (a)-(b), with total atom number values  $N=600$ in (c),(e) and  8000 in (d),(f). The quantum expectation value of the position in (c),(d) and the momentum in (e),(f) are marked by the dark solid lines, whereas the quantum uncertainties are defined by the corresponding shaded areas.} \label{Fig_px2}
\end{figure}
\begin{figure}[tbp]
\centering
\includegraphics[width=0.9\columnwidth]{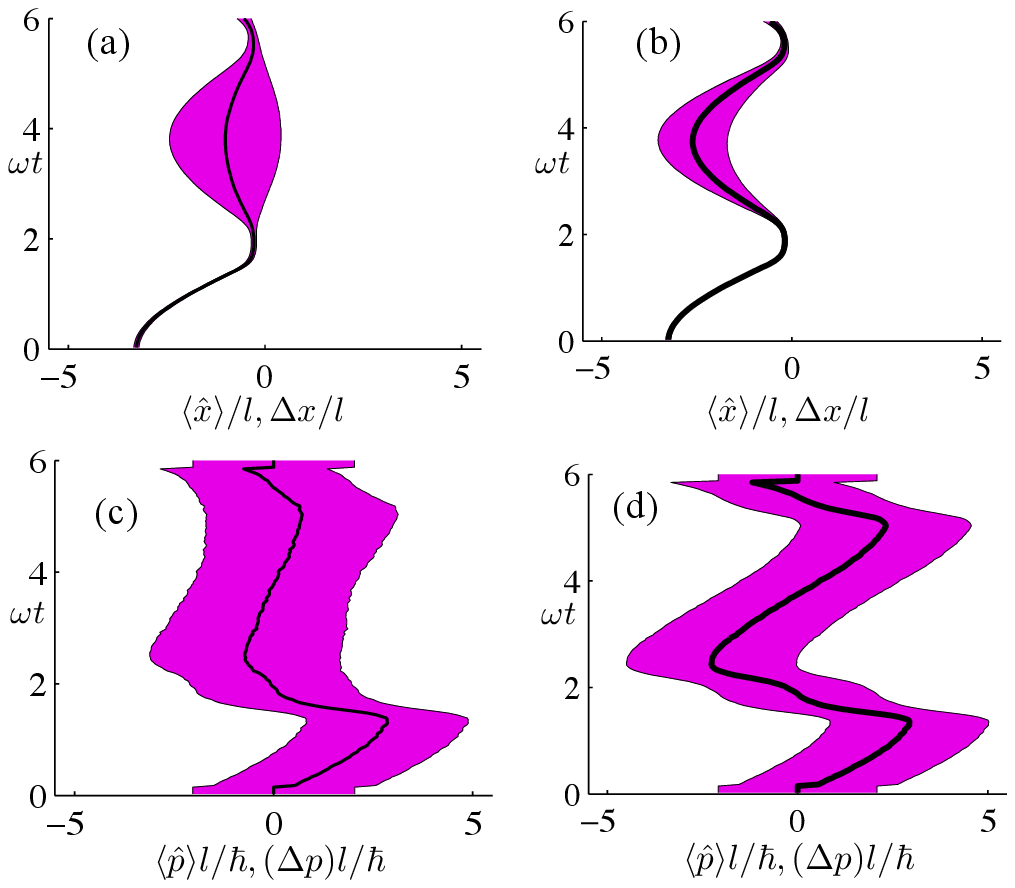}
\caption{Trajectories for the same parameters as Fig.\ \ref{Fig_px2}, but with atom numbers $N=1200$ in (a),(c) and $N=2000$ in (b),(d). The quantum expectation value of the position in (a),(b) and the momentum in (c),(d) are marked by the dark solid lines, whereas the quantum uncertainties are defined by the corresponding shaded areas.} \label{Fig_px3}
\end{figure}

In the previous sections we studied quantum effects on the relative atom number fluctuations between the solitons and on the relative phase-coherence in the soliton-barrier system. Here
we investigate the effects of quantum fluctuations on soliton position and momentum by extracting the quantum expectation values and uncertainties for each soliton's position and momentum from TWA. As described in Sec.\ \ref{Sec_TWA} we numerically track the centre-of-mass position $q_j$ and velocity $v_j$ of each soliton, as well as the atom number $N_j$ in every soliton, as a function of time in every stochastic realization. The quantum mechanical expectation values of the first and second powers of soliton position and momentum: $\langle \hat{x}_j\rangle$, $\langle \hat{x}_j^2\rangle$, $\langle \hat{p}_j \rangle$ and $\langle \hat{p}_j^2 \rangle$ may then be obtained by calculating ensemble averages of $\<  q_j\>$, $\< q_j^2\>$, $\<  \nu_j\>=\<d  q_j/dt\>$, etc., from the TWA simulations and using the relationships Eqs.\ (\ref{Eq_Avx})-(\ref{Eq_AvN}) over many realizations, so that the quantum mechanical position and momentum uncertainties for the solitons approximately satisfy
\begin{eqnarray}
(\Delta x_j)^2 & \equiv \langle \hat{x}_j^2\rangle-\langle \hat{x}_j\rangle^2 \simeq \frac{l_r^4\pi^2}{12 a^2 N_j^2} + (\Delta q_j)^2\,, \label{xwidthqm}\\
(\Delta p_j)^2 & \equiv \langle \hat{p}_j^2\rangle-\langle \hat{p}_j\rangle^2 \simeq \frac{\hbar^2 a^2 N_j^2}{3 l_r^4} + m (\Delta \nu_j)^2\,,\label{pwidthqm}
\end{eqnarray}
where $\Delta q_j=\left(\langle {q_j}^2 \rangle-\langle {q_j} \rangle^2\right)^{1/2}$ is the uncertainty of the centre-of-mass position of the soliton obtained from the stochastic TWA simulations. The actual values of the uncertainties may slightly differ from the approximate right-hand-side expressions of Eqs.~(\ref{xwidthqm}) and~(\ref{pwidthqm}) due to the deformations of the shape of the soliton wavefunction as a consequence of quantum fluctuations and nonlinear interactions. The position and momentum uncertainties therefore contain contributions from the width of the soliton's wavefunction as well as the quantum fluctuations in each soliton's centre-of-mass position and velocity. In the GPE limit, $(\Delta x_j)^2= (\Delta x_j)^2_s = l_r^4\pi^2/(12a^2 N_j^2)$ and $(\Delta p_j)^2= (\Delta p_j)^2_s = \hbar^2 a^2N_j^2/(3 l_r^4)$, as given by Eqs.~(\ref{xwidth}) and~(\ref{pwidth}). Hence, the product $\Delta p_j \Delta x_j$ is larger than the Heisenberg uncertainty limit by a factor of $\pi/3$. We find that in the TWA simulations quantum fluctuations increase this product further above the Heisenberg uncertainty limit, e.g., for $|U|=3\hbar\omega$ and for the small atom number $N=50$, the product $\Delta p_j \Delta x_j$ is approximately 6\%  larger than in the GPE limit.

We first consider the periodic dynamics of a single soliton. We investigate the effect of quantum fluctuations by varying $g/N$ for a fixed $U$, so that the initial states of the TWA simulations range from the classical limit of a negligible fraction of depleted atom number to the cases in which quantum fluctuations have a notable influence on the position and momentum uncertainties. In the quantum limit of large $|g/N|$, the quantum uncertainty in soliton position $\Delta x$ is large, and decreases with increasing $N$, reaching the GPE limit before $N=80000$ for the studied values of $U$. Surprisingly, for very large $|g/N|$, at early times in the soliton dynamics, the uncertainty in velocity is smaller than in the classical mean-field limit [illustrated in Fig. \ref{Fig_VelocityUncertainty}(a) for $|U|=3\hbar\omega $ with $N$ varying between 100 and 2000]. The initial velocity uncertainty in this case is dominated by the width of the classical soliton wavefunction solution in momentum space [i.e., the first term in Eq.\ (\ref{pwidthqm})], and at small $N$ there are relatively fewer atoms in each soliton due to quantum fluctuations, thus decreasing the nonlinear effect of interactions to broaden the velocity distribution. During the soliton evolution in the trap, the momentum uncertainty is affected by the dissipation of soliton velocities within the ensemble, and  $\Delta p_j$ become large in the quantum limit of small $N$ (when $|U|$ is constant) [see Fig.\ \ref{Fig_VelocityUncertainty}(a)-(c)]. A similar effect was found for trapped dark atomic solitons~\cite{Martin_PRL_2010,Martin_PRL_2010_b}.

Quantum fluctuations also appear to have an unexpected effect on the quantum expectation values of the soliton momentum [Fig.\ \ref{Fig_px1}(b)], with $\langle \hat{p}_j\rangle$ decreasing in the quantum limit of small $N$. For instance, at $|U|=3\hbar\omega$ and  $N=300$ ($g/N=-3.3\times 10^{-5}\hbar\omega l$), the maximum expectation value of the soliton's speed is reduced by a few percent compared with the  $N=1200$ ($g/N=-2.08\times 10^{-6}\hbar\omega l$) case of the same nonlinearity $|U|=3\hbar\omega $.
For dark solitons in tightly-confined bosonic atomic gases the speed is reduced due to phase fluctuations across the phase kink and a nonlinear relationship between the value of the phase and the speed~\cite{Martin_PRL_2010,Martin_PRL_2010_b}. In the case of a bright soliton the changes in the quantum expectation value of the momentum, however, are more likely to be an indication of dissipative centre-of-mass transport of the atoms~\cite{Fertig_PRL_2005,Isella_PRL_2005}.

Previous quantum mechanical treatments of bright solitons with very small atom numbers interacting with a barrier have found Schr\"{o}dinger-cat states of solitons, i.e., quantum superpositions of solitons traveling in both directions after hitting the barrier \cite{Cederbaum_PRL_2008, Weiss_PRL_2009}. Such superposition states cannot be represented by classical stochastic phase-space methods. In systems with large atom numbers which we consider in this work, however, any interaction with the environment and a loss of atoms from the system would rapidly decohere a macroscopic superposition state into a statistical mixture of different atom number states \cite{ZUR91,WAL85} that can be represented by TWA. In the simulations we find regimes in which the soliton propagates either left or right depending on the particular stochastic realization of $\psi_W$ even though all the physical parameters remain the same [Fig.\ \ref{Fig_px2}(a)-(b)]. We consider a barrier given by Eq.\ (\ref{Eq_Barrier}) with height $V_0\simeq6.0\hbar\omega$ and width $\sigma=0.32 l$, wider than the barriers previously considered in this paper, and approximately double the width of the barriers in the recent experiment \cite{Huletexp}.  We use soliton displacements of $x_d\simeq3.2 l$ such that the kinetic energy of the soliton as it hits the barrier is $T_E\simeq 5.2\hbar\omega$. For small values of $|U|$, the barrier splits a soliton into two in each realization of $\psi_W$. However, increasing the nonlinearity such that $|U|=15\hbar\omega$, the solitons are either entirely transmitted, or entirely reflected from run to run in different realizations of $\psi_W$. This is indicated in the increased quantum uncertainty in soliton position after the soliton hits the barrier [Fig.\ \ref{Fig_px2}(c)]. By increasing $N$, while at the same time keeping $U$ constant, we find the classical limit where the soliton is reflected in every realization of $\psi_W$ and the position and momentum uncertainties remain constant during the soliton dynamics; we find that this limit is reached at $N=80000$. The momentum uncertainty  $\Delta p_i$ is dominated by the width of the classical solution to the soliton wavefunction, rather than by the dissipation of the soliton momenta within the ensemble. However, fluctuations have a marked effect on the momentum expectation values, deforming the oscillatory behavior of $\langle p\rangle$ [see Figs \ref{Fig_px2}(e)-(f)].

\section{Concluding remarks}

We studied a physical realization of an atom-optical soliton beam-splitter and an interferometer in a harmonic trap where a repulsive laser barrier splits a soliton into two and recombines the split solitons.
We found that the classical mean-field analysis based on the GPE simulations, that does not provide any information about the atom statistics of the solitons, is not sufficient to estimate the accuracy of the interferometer even in the limit of a very weak quantum depletion and phase decoherence. This is because the nonlinear soliton-barrier interactions lead to super-Poissonian atom number fluctuations that substantially deteriorate the performance of the interferometer. The result may have implications also to other possible interferometer applications in nonlinear systems in which a high sensitivity of the interferometer output on the induced phase difference suggests a possibility for high precision measurements, emphasizing the importance of a detailed statistical analysis of the particle number fluctuations beyond the mean-field theory.

We also studied the periodic dynamics of oscillating bright soliton in the regime of considerable quantum fluctuations. Quantum fluctuations were found to have a notable effect on the splitting, dynamics and recombination of the solitons, resulting, e.g., in the condensate fragmentation and in non-classical correlations (spin-squeezing) between the solitons even for large atom numbers. The bright soliton dynamics also exhibits large quantum fluctuations of the soliton's position and velocity. At the same time the soliton can represent a truly mesoscopic quantum object that may consist of hundreds or even thousands of atoms. This provides an interesting opportunity to vary the size of the solitons and use them as an experimental probe of the quantum-classical interface \cite{Anglin_NaturePhys_2008}.

\section*{Acknowledgements}

We acknowledge financial support from the Leverhulme Trust and the EPSRC and discussions with Randall G.\ Hulet.

\end{document}